\documentclass[fleqn,usenatbib]{mnras}

\usepackage{newtxtext,newtxmath}

\usepackage[T1]{fontenc}
\usepackage{ulem} 
\usepackage{float}
\DeclareRobustCommand{\VAN}[3]{#2}
\let\VANthebibliography\thebibliography
\def\thebibliography{\DeclareRobustCommand{\VAN}[3]{##3}\VANthebibliography}

\usepackage{graphicx}	
\usepackage{amsmath}	

\newcommand{\myslash}{\operatorname{/}}
\newcommand{\myleftbracket}{\left[}
\newcommand{\myrightbracket}{\right]}

\newcommand{\teff}{\mbox{$T_{\text{eff}}$}}
\newcommand{\msun}{\mbox{M$_{\odot}$}}
\newcommand{\logg}{\mbox{$\log g$}}
\newcommand{\feh}{\mbox{[Fe/H]}}

\newcommand{\logrhk}{\mbox{$\log R^{\prime}_{\text{HK}}$}}

\newcommand{\halpha}{\mbox{H$_{\alpha}$}}

\newcommand{\dwirt}{\mbox{$\Delta W_{\textrm{IRT}}$}}
\newcommand{\dsindex}{\mbox{$\Delta S_{\textrm{HK}}$}}
\newcommand{\dfeh}{\mbox{$\Delta$[\textrm{Fe/H]}}}
\newcommand{\sigmafeh}{\mbox{$\sigma_{[\textrm{Fe/H]}}$}}
\newcommand{\sindex}{\mbox{$S_{\textrm{HK}}$}}
\newcommand{\dxh}{\mbox{$\Delta$[\textit{X}/\textrm{H}]}}
\newcommand{\dxo}{\mbox{$\Delta$[\textit{X}/\textrm{O}]}}
\newcommand{\dxhdtcont}{$\frac{\partial \Delta \myleftbracket X \myslash \rm{H} \myrightbracket}{\partial T_{\rm{cond}}}$}
\newcommand{\dxhdwirt}{$\frac{\partial \Delta \myleftbracket X \myslash \rm{H} \myrightbracket}{\partial \Delta W_{\textrm{IRT}}}$}
\newcommand\bc[1]{\textcolor{black}{\textrm{#1}}}

\title[Refractory depletion correlates with activity]{\centering{C3PO IV: co-natal stars depleted in refractories are magnetically more active--- possible imprints of planets}}

\author[Yu et al.]{Jie Yu$^{1,2},$\thanks{E-mail: jie.yu@anu.edu.au (JY)}
Yuan-Sen Ting$^{3,4,5}$,
Luca Casagrande$^{2,6}$,
Fan Liu$^{6, 7}$,
Sharon X. Wang$^{8}$,
Qinghui Sun$^{9,8}$,
\newauthor
Daniel Huber$^{10,11}$,
Boquan Chen$^{3,4}$,
Giacomo Cordoni$^{2,6}$,
Gary Da Costa$^{2,6}$,
Chelsea X. Huang$^{12}$,
\newauthor
Amanda I. Karakas$^{6, 7}$,
Shourya Khanna$^{13}$,
Junhui Liu$^{14}$,
Melissa K. Ness$^{2,6,15,16}$,
Thomas Nordlander$^{2,6,17}$,
\newauthor
John Taylor$^{1}$
\\
$^{1}$School of Computing, Australian National University, Acton, ACT 2601, Australia\\
$^{2}$Research School of Astronomy \& Astrophysics, Australian National University, Cotter Rd., Weston, ACT 2611, Australia\\
$^{3}$Department of Astronomy, The Ohio State University, 1251 Wescoe Hall Dr., Columbus, Ohio, 43210, USA\\
$^{4}$Center for Cosmology and AstroParticle Physics, The Ohio State University, 191 West Woodruff Avenue, Columbus, Ohio, 43210, USA\\
$^{5}$ Max Planck Institute for Astronomy, Königstuhl 17, D-69117 Heidelberg, Germany \\
$^{6}$ARC Centre of Excellence for All Sky Astrophysics in 3 Dimensions (ASTRO 3D), Stromlo, Australia\\
${^7}$School of Physics and Astronomy, Monash University, Melbourne, VIC, 3800, Australia\\
${^8}$Department of Astronomy, Tsinghua University, Beijing, 100084, China\\
${^9}$Tsung-Dao Lee Institute, Shanghai Jiao Tong University, Shanghai, 200240, China\\
$^{10}$Institute for Astronomy, University of Hawai‘i, 2680 Woodlawn Drive, Honolulu, HI 96822, USA\\
$^{11}$Sydney Institute for Astronomy (SIfA), School of Physics, University of Sydney, NSW 2006, Australia\\
$^{12}$University of Southern Queensland, Centre for Astrophysics, West Street, Toowoomba, QLD 4350 Australia\\
$^{13}$INAF - Osservatorio Astrofisico di Torino, via Osservatorio 20, 10025 Pino Torinese (TO), Italy\\
$^{14}$Department of Astronomy, Xiamen University, Xiamen, Fujian 361005, People's Republic of China\\
$^{15}$Department of Astronomy, Columbia University, Pupin Physics Laboratories, New York, NY 10027, USA\\
$^{16}$Center for Computational Astrophysics, Flatiron Institute, 162 Fifth Avenue, New York, NY 10010, USA\\
$^{17}$Division of Astronomy and Space Physics, Department of Physics and Astronomy, Uppsala University, Box 516, SE-75120 Uppsala, Sweden\\
}

\date{Accepted XXX. Received YYY; in original form ZZZ}

\pubyear{2015}

\begin{document}
\label{firstpage}
\pagerange{\pageref{firstpage}--\pageref{lastpage}}
\maketitle

\begin{abstract}
Chemical abundance anomalies in twin stars have recently been considered tell-tale signs of interactions between stars and planets. While such signals are prevalent, their nature remains a subject of debate. On one hand, exoplanet formation may induce chemical depletion in host stars by locking up refractory elements. On the other hand, exoplanet engulfment can result in chemical enrichment, both processes potentially producing similar differential signals. In this study, we aim to observationally disentangle these processes by using the Ca II infrared triplet to measure the magnetic activity of 125 co-moving star pairs with high SNR, high-resolution spectra from the Magellan, Keck, and VLT telescopes. We find that co-natal star pairs in which the two stars exhibit significant chemical abundance differences also show differences in their magnetic activity, with stars depleted in refractories being magnetically more active. Furthermore, the strength of this correlation between differential chemical abundances and differential magnetic activity increases with condensation temperature. One possible explanation is that the chemical anomaly signature may be linked to planet formation, wherein refractory elements are locked into planets, and the host stars become more active due to more efficient contraction during the pre-main-sequence phase or star–planet tidal and magnetic interactions.
\end{abstract}

%


\section{Introduction}

It has long been observed that compared to $\sim$85\% of solar twins, the Sun exhibits a depletion in the elemental abundance of refractory elements (such as Fe, Ti, and Al)\footnote{Following \citet{flores2023}, we categorize elements as refractory or volatile based on their condensation temperatures, with refractory elements having temperatures above 900 K and volatile elements below 900 K. The condensation temperature values are taken from \citet{lodders2003}.}, and this depletion is correlated with the condensation temperature \citep[e.g.,][]{melendez2009, bedell2018, Rampalli2024}. Similar phenomena have also been observed and validated in other stars \citep[e.g.][]{ramirez2010, ramirez2014, liu2020, flores2023, liu2024}.

The connection between this phenomenon and planets has been investigated by the following studies. For example, \citet{melendez2009} and \citet{ramirez2009} proposed that the sequestration of these refractory elements by terrestrial planets may explain this depletion. Additionally, \citet{booth2020} suggested that giant planets, such as Jupiter, could contribute by creating dust traps, thereby preventing the fall of refractory-rich dust on the host star. Recently, \citet{huhn2023} conducted numerical simulations on the abundance differences resulting from such dust traps and found that the measurements of the elemental abundance differences for a wide binary system, which hosts a giant planet, align reasonably well with their simulations.

Unlike chemical depletion that may be caused by exoplanet formation processes, chemical enrichment can result from the ingestion of planetary material \citep[e.g.,][]{chambers2010, huhn2023}. Based on this hypothesis, \citet{liu2024} reported evidence of planetary ingestion in at least 8\% of stars, a rate comparable to observations in solar twins \citep{ramirez2009} and N-body simulations \citep{bitsch2023}. This conclusion is supported by the consistency between observed elemental abundance differences and theoretical models of planetary engulfment.

In this work, we revisit the data set as in \citet{liu2024}, but focusing on magnetic activity, which is directly derived from the spectra. Magnetic activity is closely linked to stellar rotation, with stars exhibiting increasingly stronger activity as rotation rates increase. This activity reaches saturation at rapid rotation, characterized by rotation periods of less than $\sim$3 days \citep[see][for a recent review]{isik2023}. In close binaries, due to tidal interactions, stars can spin more rapidly when their rotation becomes synchronized with orbital periods. Consequently, these binary stars, even at the old ages, may exhibit enhanced activity, as seen in red giants \citep[e.g.,][]{gehan2022}. Therefore, it is likely that hot Jupiters and their hosts may undergo analogous interactions.

\begin{figure}
\begin{center}
\resizebox{\columnwidth}{!}{\includegraphics{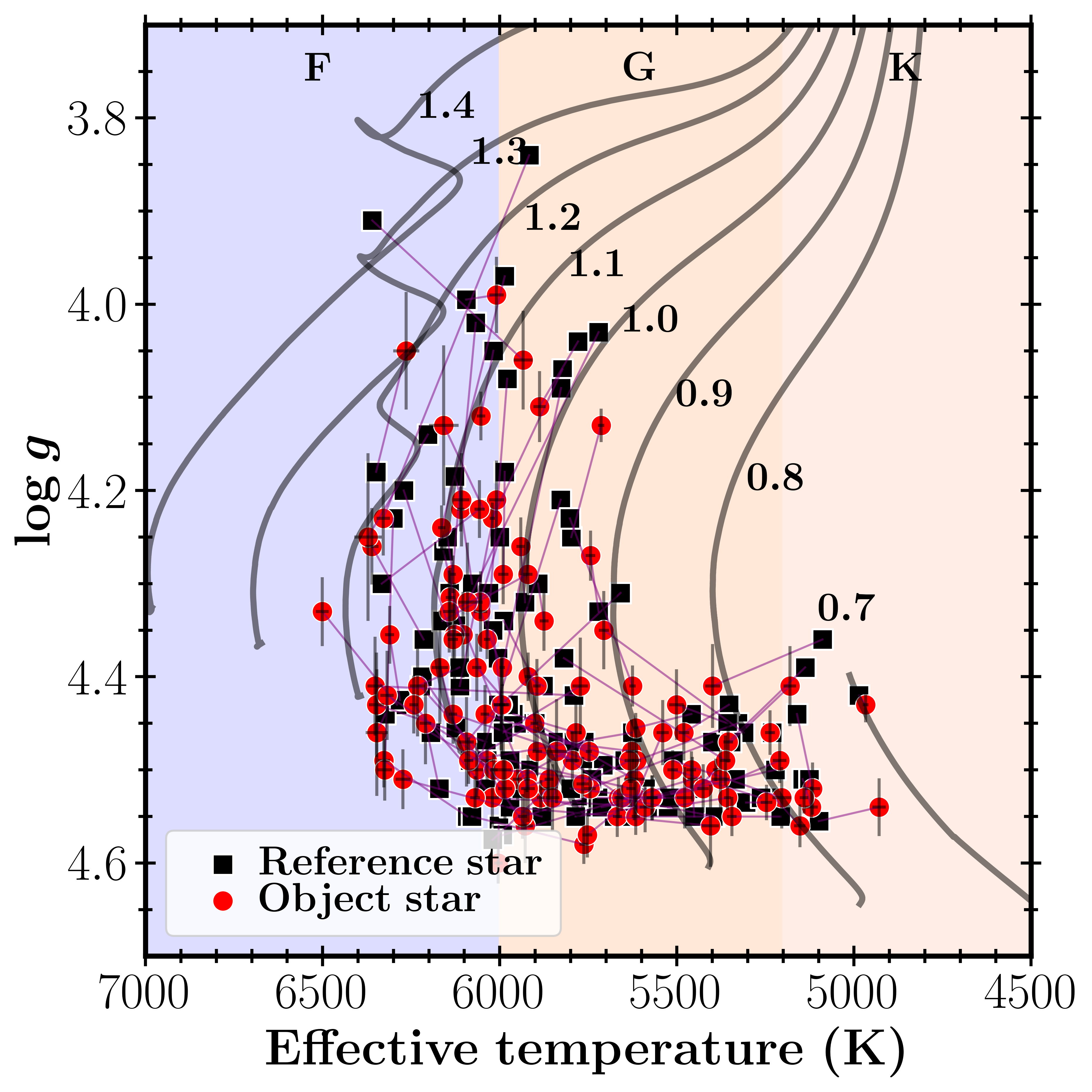}}
\caption{Kiel diagram of 125 co-moving systems, each comprising a reference star (black square) and an object star (red circle), connected by a purple line. The designation of reference and object stars is arbitrary in this work but does not affect our conclusions if reversed. Grey curves represent \texttt{MIST} evolutionary tracks for solar-metallicity stars with masses ranging from 0.7 to 1.4~\msun\ in 0.1~\msun\ increments. The \teff\ and \logg\ values are adopted from \citet{liu2024}. The background is shaded with three colors, approximating the true colors of F, G, and K stars as they would appear if viewed from space \citep{harre2021}.}
\label{fig:kiel}
\end{center}
\end{figure}

Tidal interactions between planets and their host stars have been extensively studied in the literature \citep[see recent reviews by][]{ogilvie2020, lanza2022}, and these interactions manifest primarily in two mechanisms: equilibrium tides and dynamical tides \citep[e.g.,][]{ogilvie2004, mathis2015}. Equilibrium tides result from the lag between the tidal bulge raised on the star by the planet and the planet’s position in its orbit. This lag transfers angular momentum between the orbit and the star’s spin, altering the star’s rotation over time \citep{ogilvie2014}. Complementing equilibrium tides are dynamical tides, which excite stellar oscillations through the gravitational potential of the planet and the Coriolis force of the star \citep{ogilvie2007}. The efficiency of dynamical tides depends on stellar properties, particularly the thickness of the outer convective zone \citep{mathis2015}.

In accordance with theoretical predictions on tides, observations by \citet{pont2009} and \citet{brown2011} revealed that certain host stars of hot Jupiters exhibit unexpectedly rapid rotation rates. Additionally, \citet{poppenhaeger2014} observed that host stars expected to experience strong tidal interactions with their planets display enhanced X-ray activity compared to their counterparts in wide binary systems devoid of known planets. Furthermore, direct instances of magnetic star–planet interactions have been found, where stars hosting short-period planets exhibit significantly higher flare rates \citep[e.g.,][]{kavanagh2021, feinstein2024}, or excess flaring aligned with the planet's phase or close to periastron \citep[e.g.,][]{shkolnik2005, pillitteri2011, maggio2015, ilin2023}.

In addition to tidal and magnetic interactions, giant planets can further shape the rotational and magnetic evolution of their host stars. As \citet{booth2020} suggested, the formation of dust gaps can magnetically decouple the host star from the planet, enabling more efficient stellar contraction during the pre-main-sequence (PMS) phase. This process shortens the lifetime of the disk and causes the host star to spin up more rapidly. As a result, these stars exhibit faster rotation rates and stronger magnetic activity during the main sequence \citep[e.g.,][]{monsch2023}. Thus, the presence of giant planets may influence not only stellar chemistry but also the rotation and magnetic activity of their host stars. However, few studies have investigated the connection between magnetic activity and the observed chemical anomalies in these stars. The motivation for this work is to observationally explore whether magnetic activity can help disentangle the closely related differential signals seen in stellar chemical anomalies.

\begin{figure}
\begin{center}
\resizebox{\columnwidth}{!}{\includegraphics{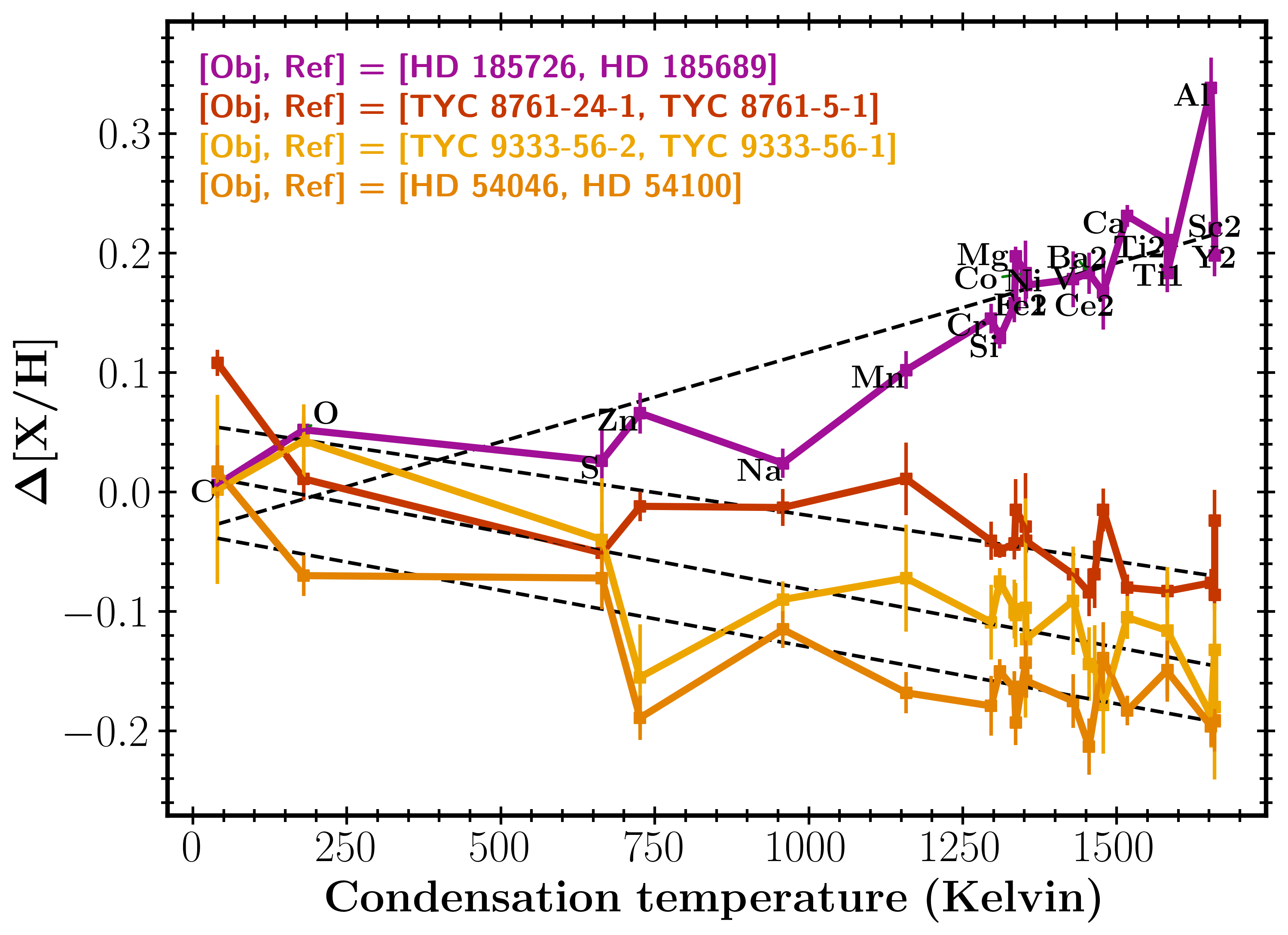}}
\caption{Four example co-moving systems illustrating the relationship between condensation temperature and elemental abundance differences for 21 elements, each labeled in the plot. The measurements of elemental abundance differences are based on high-resolution spectra collected through the C3PO program. The vertical axis shows the elemental abundance differences between the two stars in each pair. The assignment of reference and object stars is arbitrary; however, it does not change the main conclusions of this work. The dashed lines represent the best linear fit for each system, with star names indicated in the top left corner and color-coded to match the corresponding curves.}
\label{fig:xhtcond}
\end{center}
\end{figure}

\begin{figure*}
\begin{center}
\resizebox{0.7\textwidth}{!}{\includegraphics{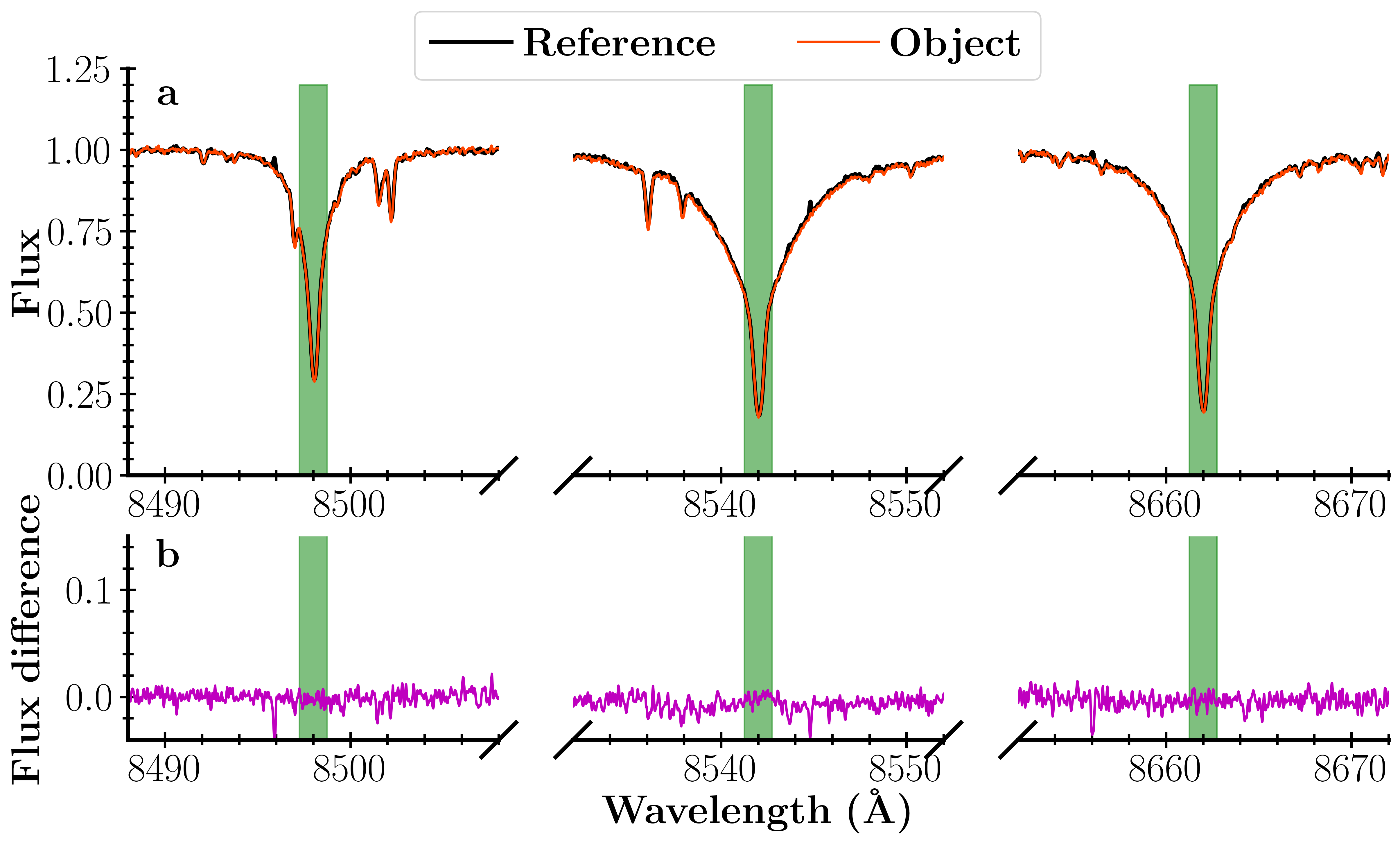}}\\
\resizebox{0.7\textwidth}{!}{\includegraphics{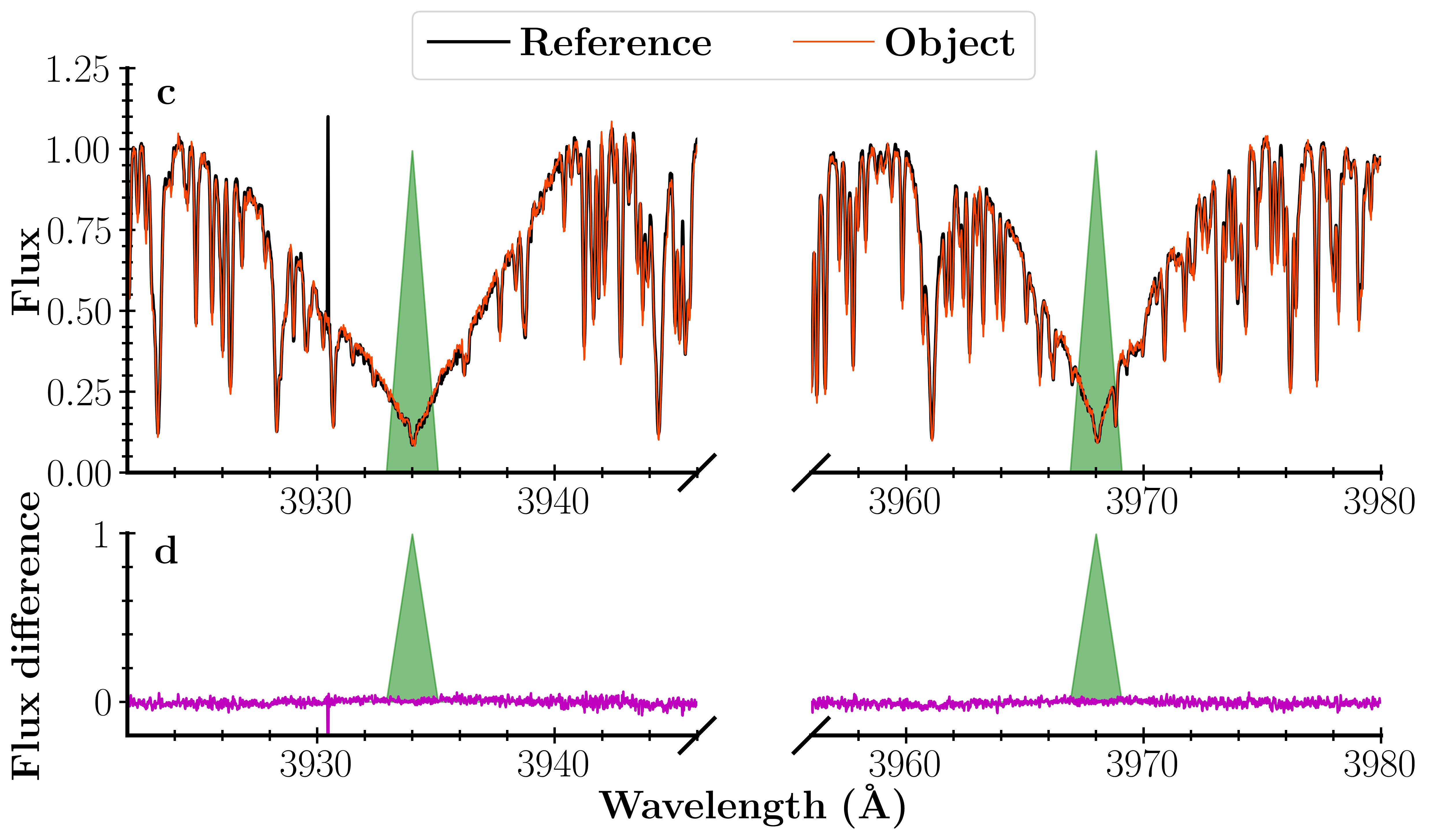}}
\caption{Example of flux-normalized spectra in the rest frame for a co-moving stellar pair (TYC 8702-53-1/TYC 8702-170-1) with negligible differential activity. \textbf{Panel a}: The Ca II infrared triplet lines (8498 Å, 8542 Å, and 8662 Å) for the reference star (black) and object star (red). The shaded green region represents a 0.75 Å window centered on each triplet line, used to calculate the differential activity index. \textbf{Panel b}: The flux difference between the object and reference stars. \textbf{Panels c} and \textbf{d}: Similar to the top two panels, but for the Ca II H \& K lines (3934 Å and 3968 Å). A triangular filter with a full width at half maximum (FWHM) of 1.09 Å is applied, as shown in \textbf{panels c} and \textbf{d}, to integrate the emission linked to stellar activity.}
\label{fig:ExampleWeakActivity}
\end{center}
\end{figure*}

\section{Data and Methods}
\subsection{Data and Sample Selection}\label{sample}
Understanding the relationship between stellar chemical anomalies and planetary processes requires careful analysis of stars that share similar initial conditions. Co-moving and spatially close stellar pairs, often classified as co-natal, provide a valuable opportunity for such studies \citep{kamdar2019, nelson2021}. Co-natal stars are thought to have originated from the same molecular cloud, implying a shared chemical history and evolutionary pathway. By comparing differential elemental abundances in these stars, we can isolate variations that may arise from external factors such as ISM inhomogeneity and atomic diffusion. This investigation forms the basis of the \textbf{C}omplete {\bf C}ensus of \textbf{C}o-moving \textbf{P}airs \textbf{O}f stars (C3PO) program, which is dedicated to high-precision, homogeneous differential elemental abundance analysis of co-moving star pairs \citep{yong2023, liu2024, sun2025}.

\begin{figure*}
\begin{center}
\resizebox{0.7\textwidth}{!}{\includegraphics{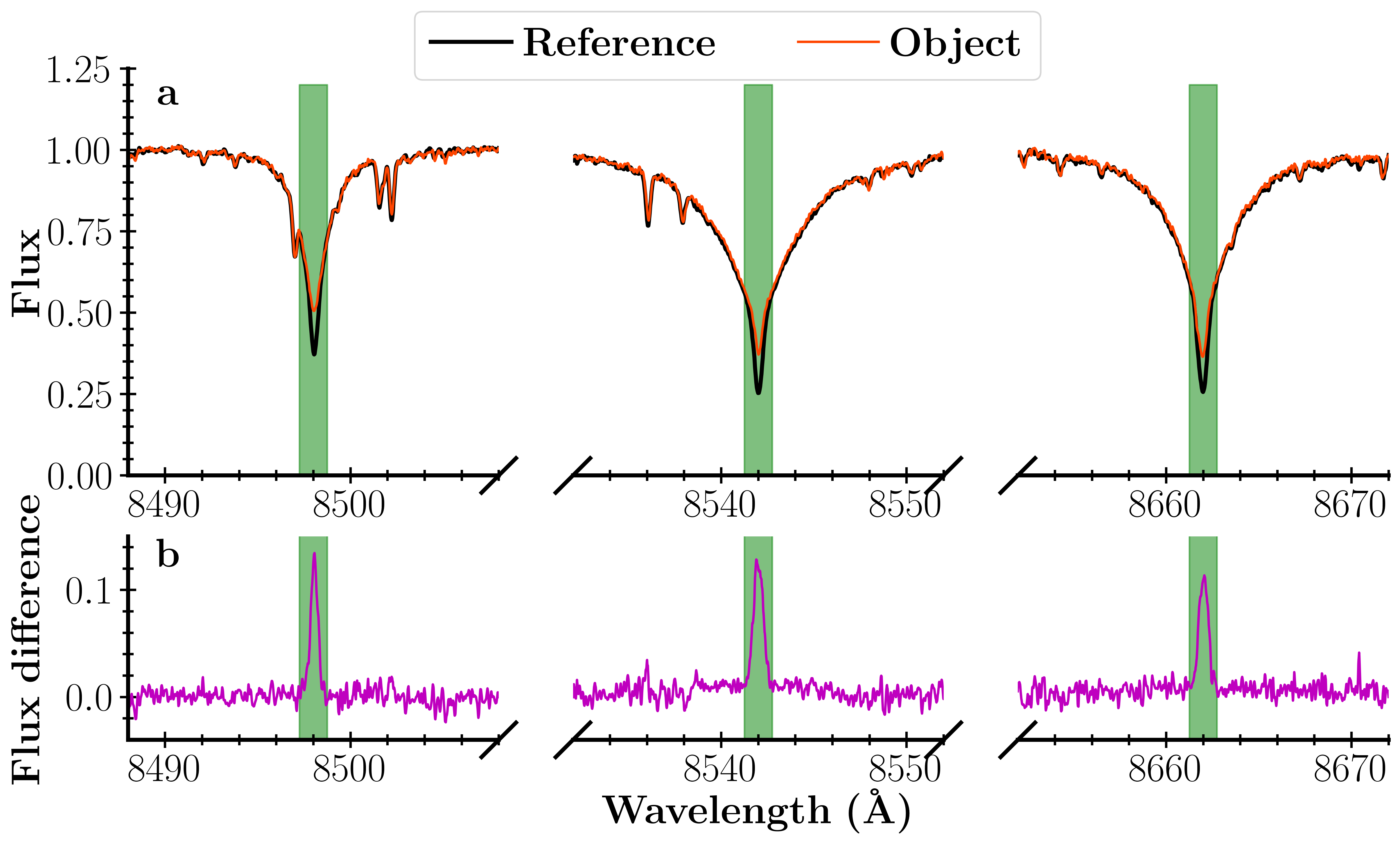}}\\
\resizebox{0.7\textwidth}{!}{\includegraphics{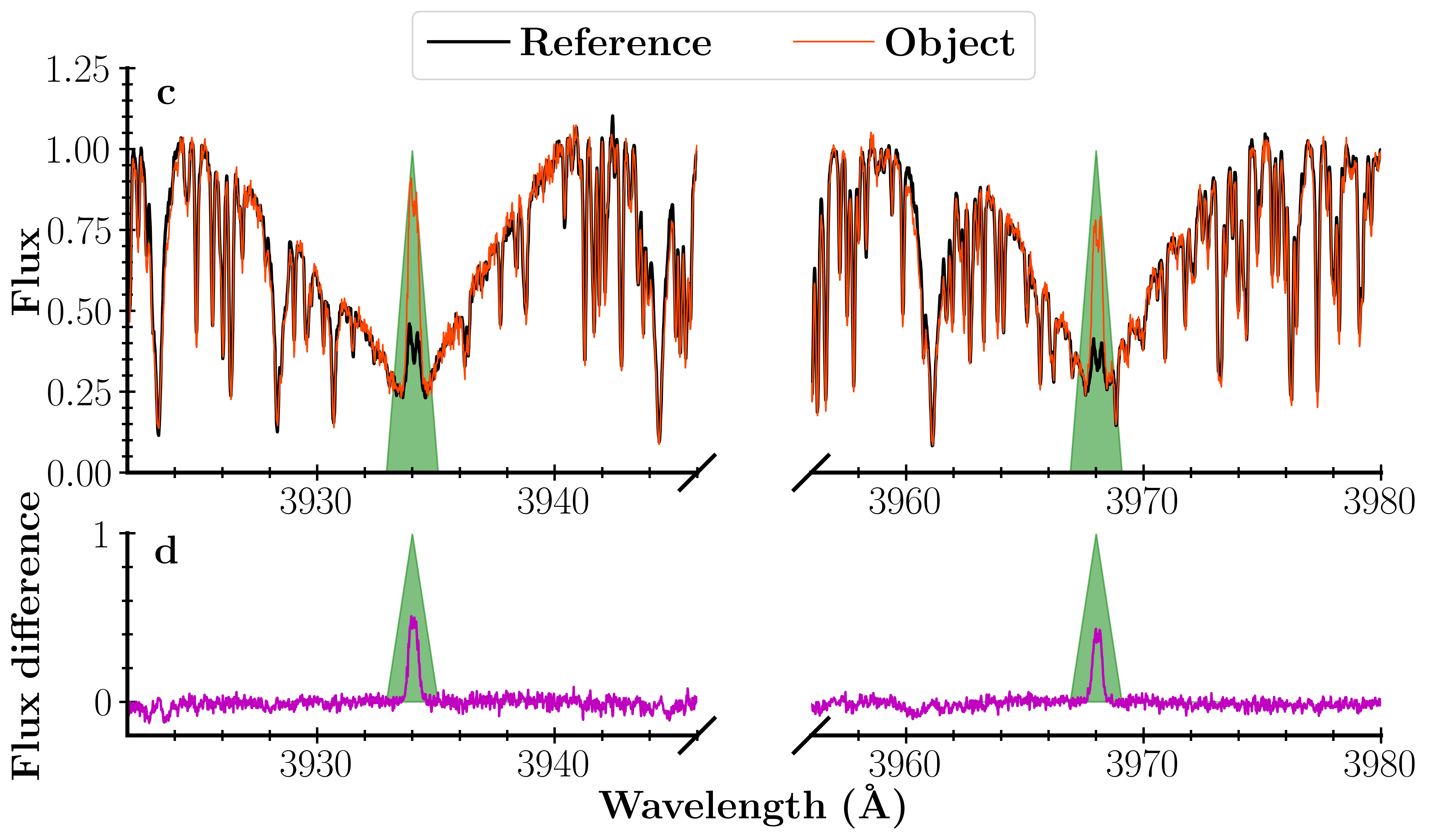}}
\caption{Similar to Fig.~\ref{fig:ExampleWeakActivity}, except for another co-moving system (TYC 8832-1422-1/TYC 8832-243-1) having strong differential activity in both Ca II infrared triplets (\textbf{panels a} and \textbf{b}) and Ca II H~\&~K lines (\textbf{panels c} and \textbf{d}).}
\label{fig:ExapleStrongActivity}
\end{center}
\end{figure*}

The C3PO program collected high-resolution, high signal-to-noise ratio (S/N $\approx$ 250 per pixel at 600 nm) spectra using three major instruments: the MIKE spectrograph on the Magellan Telescope over 7 nights ($R \approx 50,000$), HIRES on the Keck Telescope over 1 night ($R \approx 72,000$), and UVES on ESO’s Very Large Telescope over 26.4 hours ($R \approx 110,000$). Each pair of stars in our sample has similar colour ($|\Delta(BP - RP)| \leq 0.15$) and absolute magnitude ($|\Delta M_G| < 1$) in the Gaia G band. This similarity between pairs facilitated the line-by-line differential analysis, a crucial approach for achieving high precision in stellar parameters and relative elemental abundances \citep[e.g.][]{melendez2009, sun2020a}. Additionally, this method allows us to cancel out major sources of systematic uncertainties, such as those arising from model atmospheres and atomic line data \citep{bedell2014, nissen2018}.

For this study, we revisit 125 co-moving star pairs selected from the C3PO program \citep{yong2023, liu2024, sun2025}. Each pair consists of a reference star and an object star, both of which exhibit similar fundamental properties. The majority of the pairs show differences of $\Delta \teff\ < 300$K, $\Delta \logg\ < 0.3$, and $\Delta$\feh $< 0.3$. The distinction between reference and object stars is arbitrary, meaning the reference star can have either higher or lower elemental abundances than the object star. Figure \ref{fig:kiel} illustrates the distribution of these stars on the Kiel diagram, with each pair connected by a purple line, demonstrating their proximity in stellar parameters. Most of the stars lie along the main sequence, indicating that this sample represents a relatively young stellar population. The median ages, determined through isochrone fitting, are 3.25 Gyr for the object stars and 3.50 Gyr for the reference stars. The median absolute age difference between the object and reference stars is 1.30 Gyr, which is comparable to the median age uncertainty (1.68 Gyr), suggesting that their ages are equal to within uncertainties and that the paired stars are consistent with being co-natal. These ages were derived using the \texttt{isoclassify} pipeline \citep{huber2017, berger2020}, based on \teff, \logg, and \feh\ measurements, along with their respective uncertainties.

Following the methodology of \citet{liu2024}, we classify 91 of the 125 co-moving systems as co-natal, based on 3D separations $\Delta s< 10^6$ A.U. and 3D velocity differences $\Delta v < 2.0$ km/s. Among these co-natal systems, 65 exhibit homogeneous chemistry, defined as $|\dfeh| < 3\sigma_{\dfeh}$. The remaining 26 systems show chemical anomalies, characterized by $|\dfeh| \geq 3\sigma_{\dfeh}$. Here, \dfeh\ represents the metallicity difference between the two stars in each pair, and $\sigma_{\dfeh}$ corresponds to the uncertainty in \dfeh.

We adopt stellar parameters and elemental abundances from \citet{liu2024}, which were derived using the high-precision spectra described above. The typical uncertainties in the differential stellar parameters are around 15 K for \teff, 0.035 for \logg, and 0.012 dex for \feh. Based on these differential stellar parameters, elemental abundances for 21 elements (C, O, Na, Mg, Al, Si, S, Ca, Sc, Ti, V, Cr, Mn, Fe, Co, Ni, Zn, Sr, Y, Ba, and Ce) were calculated on a line-by-line basis. The average errors in differential abundances for nearly all co-moving pairs are approximately 0.015 dex (3.5\%).

In Fig.~\ref{fig:xhtcond}, we present four co-moving systems that exhibit strong correlations between elemental abundance difference (\dxh) and condensation temperature. The slopes of these correlations vary between positive and negative due to the arbitrary designation of reference and object stars in each pair. These correlations align with previous observations for several binaries of stellar twins hosting planets or debris disks \citep[see][for a review]{melendez2016}. In the next section, we introduce a differential activity index (i.e. the activity difference between the two stars in each pair) as an additional diagnostic to further investigate these chemical anomalies of these co-natal stars.

\subsection{Activity Index}\label{med: activityindex} 
We employ the residual equivalent width of the Ca II infrared triplet lines as a chromospheric activity indicator in this study, denoted as \dwirt. This metric is defined as the sum of the flux differences between the flux-normalized spectra of the paired stars, calculated as the flux of the object star minus that of the reference star, consistent with the ordering used in computing \dxh. The summation is evaluated over a narrow window, 0.75 Å wide, centered around each of the Ca II infrared triplet lines (8498~Å, 8542~Å, and 8662~Å). \bc{This \dwirt\ metric has been validated in the literature, where it has demonstrated a strong correlation with the chromospheric activity diagnostic \logrhk, derived from the Ca II H \& K lines \citep{busa2007,martin2017}. It has also been widely used to study stellar activity in large-scale surveys such as RAVE \citep{zerjal2013}, Gaia RVS \citep{lanzafame2023}, and LAMOST \citep{huang2024}.} Figs.~\ref{fig:ExampleWeakActivity} and \ref{fig:ExapleStrongActivity} show examples of the spectra and the integration windows for two star pairs---one with minimal difference and one with a pronounced difference in their line cores.

There are 82 systems with spectra covering all three triplet lines, 19 systems with spectra covering only the first two lines, and 24 systems that lack spectral coverage of the triplet lines entirely. To account for this variation, we sum the flux differences within the windows and then calculate the average by dividing the total flux difference by the number of lines measured for each star. Importantly, we verified that the number of lines used for calculating \dwirt\ does not affect our final conclusions, as confirmed by tests using a single, randomly chosen Ca II infrared triplet line. The window width used in our analysis, 0.75 Å, is half of that employed by \citet{lanzafame2023}, chosen to minimise the influence of line wings, which are unrelated to stellar activity. However, testing with a wider window of 1.5 Å showed no significant impact on our results. All spectra are shifted to the rest frame using radial velocities derived from this study by fitting a composite model, which consists of a line and a Gaussian for each absorption feature. We find this method to be particularly effective for precisely identifying local features, such as the Ca II triplet lines, and it differs from the commonly used cross-correlation function (CCF) methods, which are typically applied over broader wavelength ranges. Initial radial velocity estimates were taken from \citet{liu2024} for the fitting process.

\begin{figure}
\includegraphics[width=\columnwidth]{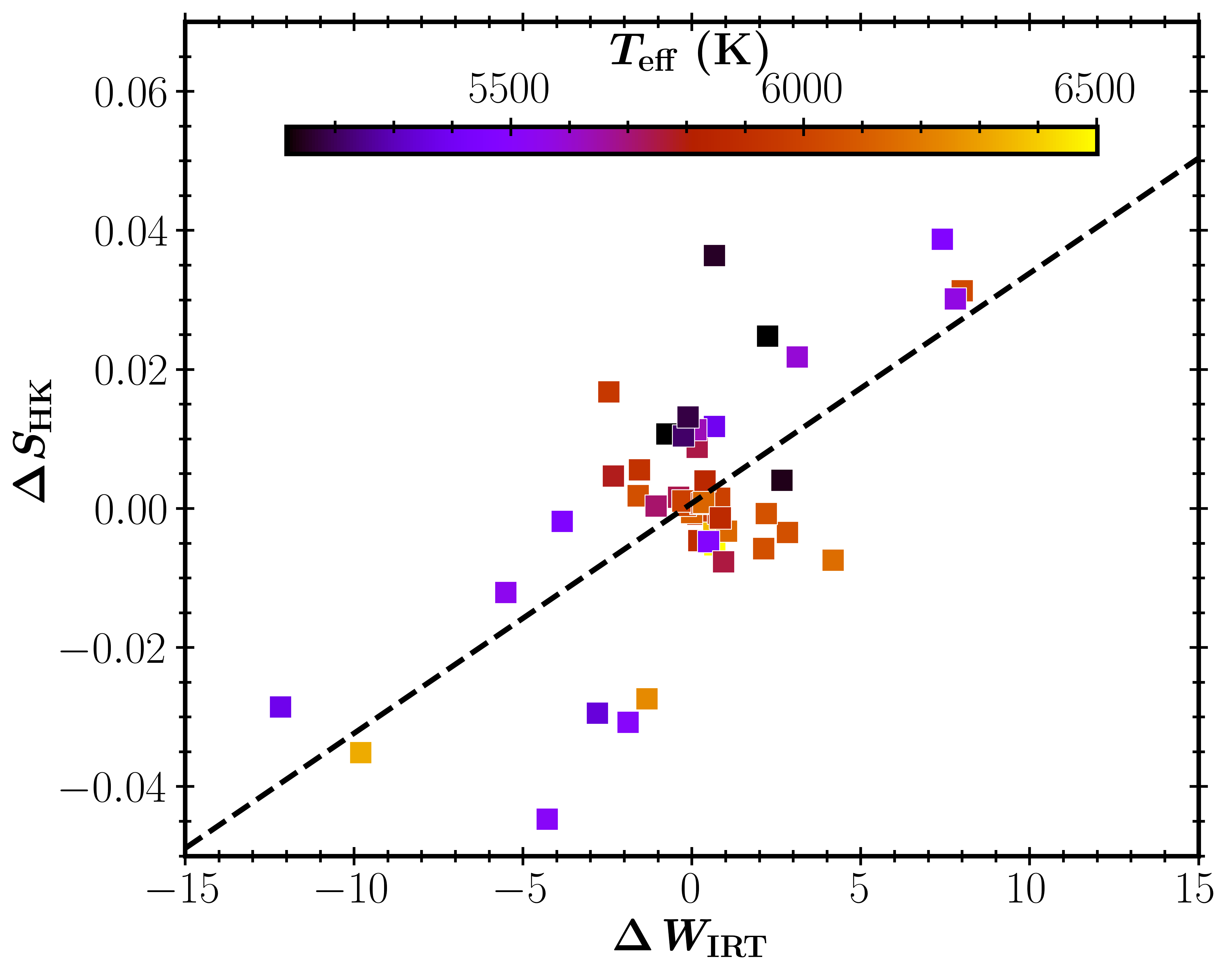}
\caption{Correlation between the differential activity index, \dwirt, obtained from the Ca II infrared triplet, and the differential activity indicator, \dsindex, derived from the Ca II H \& K lines (see Sect.~\ref{med: activityindex} ). \bc{The typical uncertainties of \dwirt\ (median=0.09) are smaller than the symbol size, primarily due to the high S/N of the C3PO spectra, and are therefore not visible}. The colour bar indicates the \teff\ of 50 object stars, whose spectra cover the two line features. The star pairs shown in the figure include those with and without chemical anomalies. 
}\label{fig:SIndex}
\end{figure}

\begin{figure}
\begin{center}
\resizebox{\columnwidth}{!}{\includegraphics{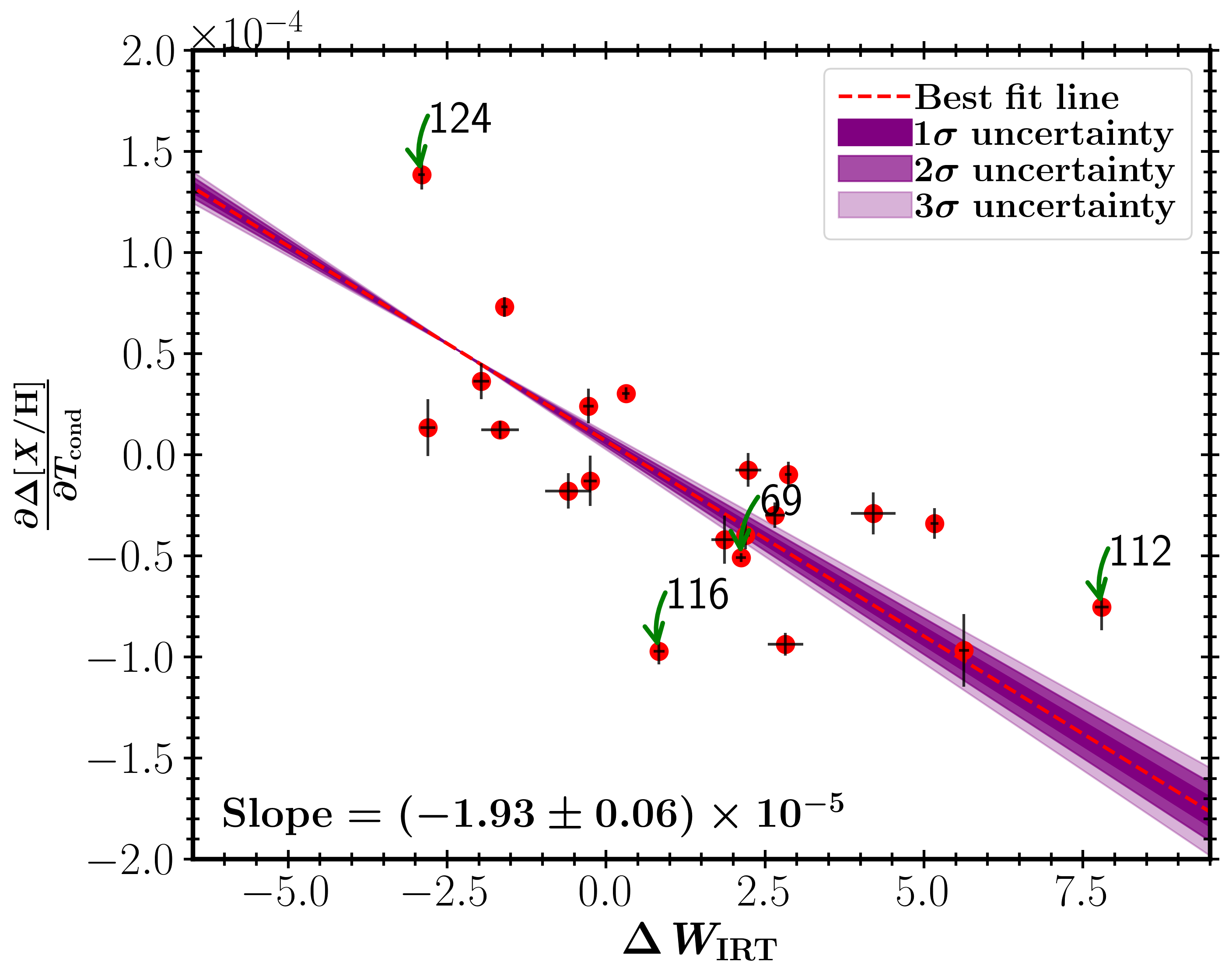}}
\caption{Differential activity index (\dwirt) as a function of the slope of the condensation temperature versus elemental abundance difference relation (\dxhdtcont) for 21 co-natal systems exhibiting chemical anomalies (i.e., $|\dfeh| \geq 3\sigma_{\dfeh}$). The red dashed line represents the best fit obtained through weighted linear regression, minimizing the weighted sum of squared residuals between observed data and the fitted line, with weights as the inverse square of the uncertainty in \dxhdtcont\ as determined from the linear fit. The shaded purple regions indicate the 1-, 2-, and 3-$\sigma$ uncertainties around the best fit, with the slope value annotated. Green arrows highlight four pairs suggested by \citet{liu2024} as potential exoplanet engulfment instances, with corresponding pair IDs.}
\label{fig:chemistrytcondactivity}
\end{center}
\end{figure}

\bc{To estimate the uncertainty in \dwirt, we propagate the flux errors in the two spectra  for the reference and object stars. Given two spectra with flux values $f_1(\lambda)$ and $f_2(\lambda)$ at each wavelength $\lambda$, the flux difference is defined as:
\begin{equation}
\Delta f(\lambda) = f_1(\lambda) - f_2(\lambda).
\end{equation}
The corresponding uncertainty in $\Delta f(\lambda)$ follows standard error propagation:
\begin{equation}
\sigma_{\Delta f}(\lambda) = \sqrt{\sigma_{f_1}^2(\lambda) + \sigma_{f_2}^2(\lambda)},
\end{equation}
where $\sigma_{f_1}(\lambda)$ and $\sigma_{f_2}(\lambda)$ are the respective flux uncertainties of the two spectra. Given that flux uncertainties are not available for the C3PO spectra \citep{yong2023}, we estimate the signal-to-noise ratio (S/N) by selecting a continuum region near each line of the Ca II infrared triplet (8474--8490\AA, 8518--8532\AA, 8626--8646\AA). Within each of these three regions, we compute the median flux \( F_{\mathrm{cont}} \) and the 1$\sigma$ scatter of the flux from the 16th and 84th percentiles of the flux distribution. The S/N is then estimated by \( F_{\mathrm{cont}} / \sigma \). With S/N, the flux uncertainties can be expressed as:
\begin{equation}
\sigma_{f_1}(\lambda) = \frac{f_1(\lambda)}{\mathrm{(S/N)}_1}, \quad \sigma_{f_2}(\lambda) = \frac{f_2(\lambda)}{\mathrm{(S/N)}_2},
\end{equation}
where $\mathrm{(S/N)}_1$ and $\mathrm{(S/N)}_2$ are the signal-to-noise ratios of the two spectra.
Since \dwirt\ is defined as the sum of flux differences over a given wavelength range:
\begin{equation}
\dwirt = \sum_{\lambda} \Delta f(\lambda),
\end{equation}
the total uncertainty in $\Delta W$ is given by:
\begin{equation}
\sigma_{\dwirt} = \sqrt{\sum_{\lambda} \sigma_{\Delta f}^2(\lambda)}.
\end{equation}
Substituting $\sigma_{\Delta f}(\lambda)$, we obtain:
\begin{equation}
\sigma_{\dwirt} = \sqrt{\sum_{\lambda} ( (\frac{f_1(\lambda)}{\mathrm{(S/N)}_1})^2 + (\frac{f_2(\lambda)}{\mathrm{(S/N)}_2})^2 )}.
\end{equation}
Similar to the calculation of \dwirt, when multiple lines in the triplet are observationally available, we compute the final uncertainty $\sigma_{\dwirt}$ by averaging over all \( N \) lines, namely, 
\begin{equation}
\sigma_{\dwirt} = \sqrt{\frac{\sum_{i=1}^{N} \sigma_{\dwirt, i}^2}{N}}
\end{equation}
}

To validate our \dwirt\ measurements, we calculate another commonly used chromospheric activity index, \sindex, based on the Ca ~II~H~\&~K lines (3968 Å and 3934 Å, respectively), following standard methodology \citep[see, e.g.,][]{zhang2020, gehan2022, yu2024,cordoni2024}. It is important to note that \dsindex\ is used solely for validation purposes and \dwirt\ serves as the primary metric for our subsequent analysis. This choice is due to wavelength coverage limitations in our spectra as determined by the observation settings---Magellan/MIKE (3350–9300 Å), Keck/HIRES (4200–8500 Å), and VLT/UVES (4800–6800 Å, whose spectra were thus not used for measuring activity indexes) \citep[see][]{yong2023}. With this spectral coverage, we measure \(\dsindex\) for 40\% of the systems and and \dwirt\ for 81\% of the star pairs.

To calculate \dsindex, we first identify the Ca II H \& K lines by redetermining the radial velocities using the \halpha\ line ($\lambda_0 = 6563$ Å in air), applying the same method used for identifying the Ca II triplet lines. We calculate the integrated emission line fluxes within the H and K bandpasses, using 1.09 Å FWHM triangular windows. The continuum fluxes within the R and V bandpasses (centered at 4001 Å and 3901 Å, respectively) are summed using 20 Å rectangular windows. We then calculate \sindex\ as follows \citep{karoff2016}:
\begin{equation} 
\sindex = \alpha \times 8 \times \frac{1.09~\text{\r{A}}}{20~\text{\r{A}}} \times \frac{H+K}{R+V}, 
\end{equation}
where the \textit{S}-index calibration factor is $\alpha = 1.8$ \citep{hall2007}. The \sindex\ measurements are then used to calculate the differential activity index, \dsindex, which is the difference in \sindex\ between the object and reference stars.

Fig.~\ref{fig:SIndex} demonstrates a clear correlation between our two differential activity indices (\dwirt\ and \dsindex) across the full temperature range of our sample. \bc{We note that the typical uncertainties in \dwirt\ (median=0.09) are smaller than the symbol size---primarily due to the high S/N of the C3PO spectra---indicating that additional factors contribute to the scatter in the correlations. One possible source of this scatter is the variation in formation regions within the chromosphere. Indeed, similar levels of scatter in activity indices also based on the Ca II H \& K and infrared triplet lines have been reported in previous studies \citep[e.g.,][]{lanzafame2023}. Nevertheless, }this correlation validates the use of \dwirt\ as a reliable measure of stellar activity. Moreover, \dwirt\ measurements primarily reflect genuine activity differences between paired stars and are minimally affected by light contamination from close companions. This is supported by our ongoing LCOGT monitoring campaign of 14 randomly selected C3PO pairs, which has revealed only one potential unseen companion among all monitored stars.

We acknowledge that one possible systematic error in measuring differential activity could arise from variations in calcium abundance ([Ca/H]) between paired stars \citep{carrera2007, dacosta2016}. Noteworthy is that [Ca/H] reported in \cite{liu2024} have been obtained using lines other than Ca II  infrared triplet or the Ca II H \& K discussed here. For instance, in Fig.~\ref{fig:ExapleStrongActivity}, the object star (red curves) has a measured [Ca/H] value 0.08 dex (18\%) lower than its reference star (black curves), which could potentially result in shallower triplet lines. However, we find this effect to be negligible based on two key observations. First, the emission features in the Ca II H \& K line cores show substantial differences between the paired stars (Fig.~\ref{fig:ExapleStrongActivity}c), with core flux variations reaching a factor of two—far too large to be explained by the small 0.08 dex difference in [Ca/H]. Second, the strong correlation between \dwirt\ and \dsindex\ across our sample (Fig.~\ref{fig:SIndex}) confirms that the observed variations primarily reflect differences in magnetic activity rather than calcium abundance.

\begin{figure}
\begin{center}
\resizebox{\columnwidth}{!}{\includegraphics[trim=0cm 1.96cm 0cm 0cm, clip]{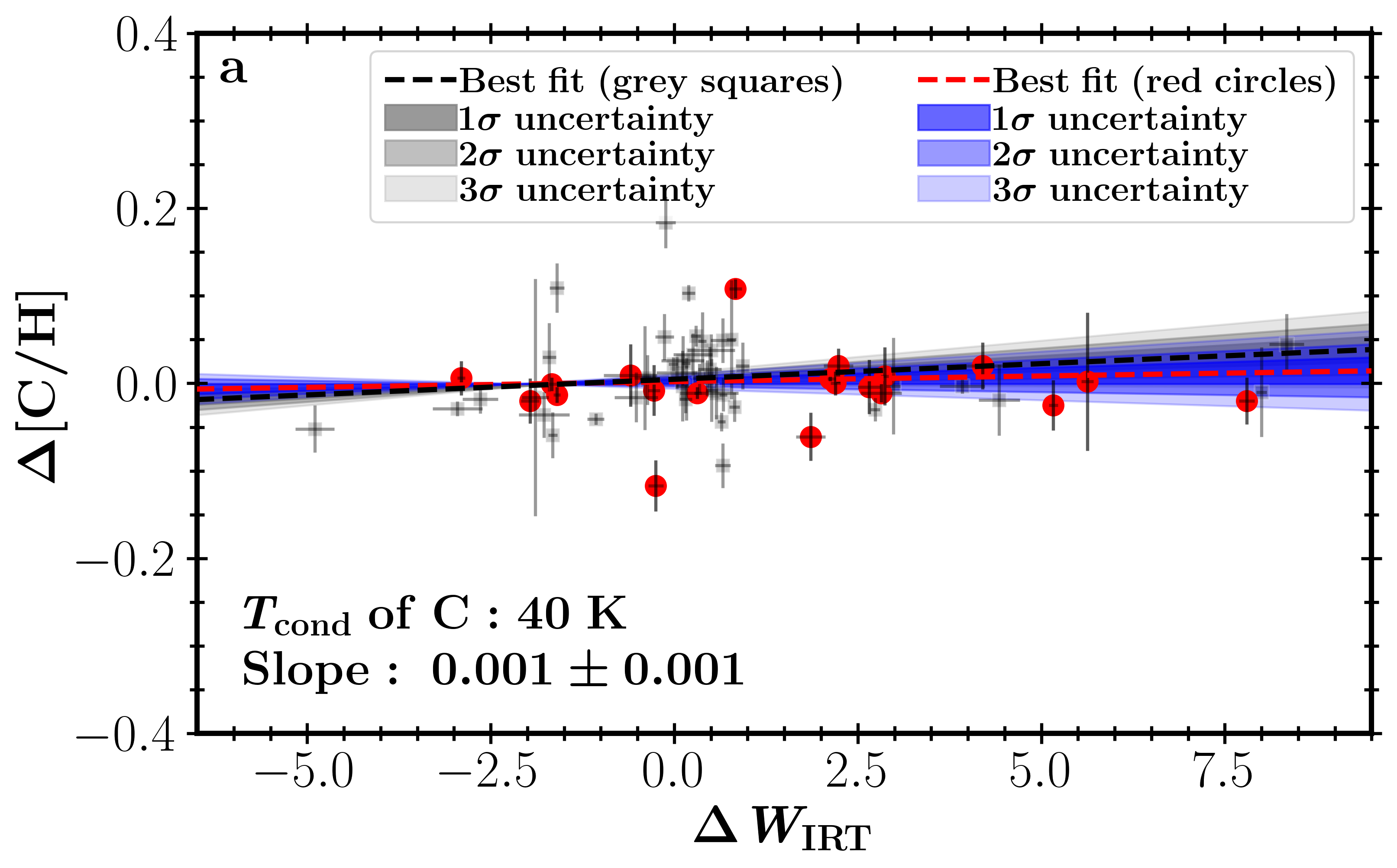}}\\
\resizebox{\columnwidth}{!}{\includegraphics[trim=0cm 1.96cm 0cm 0cm, clip]{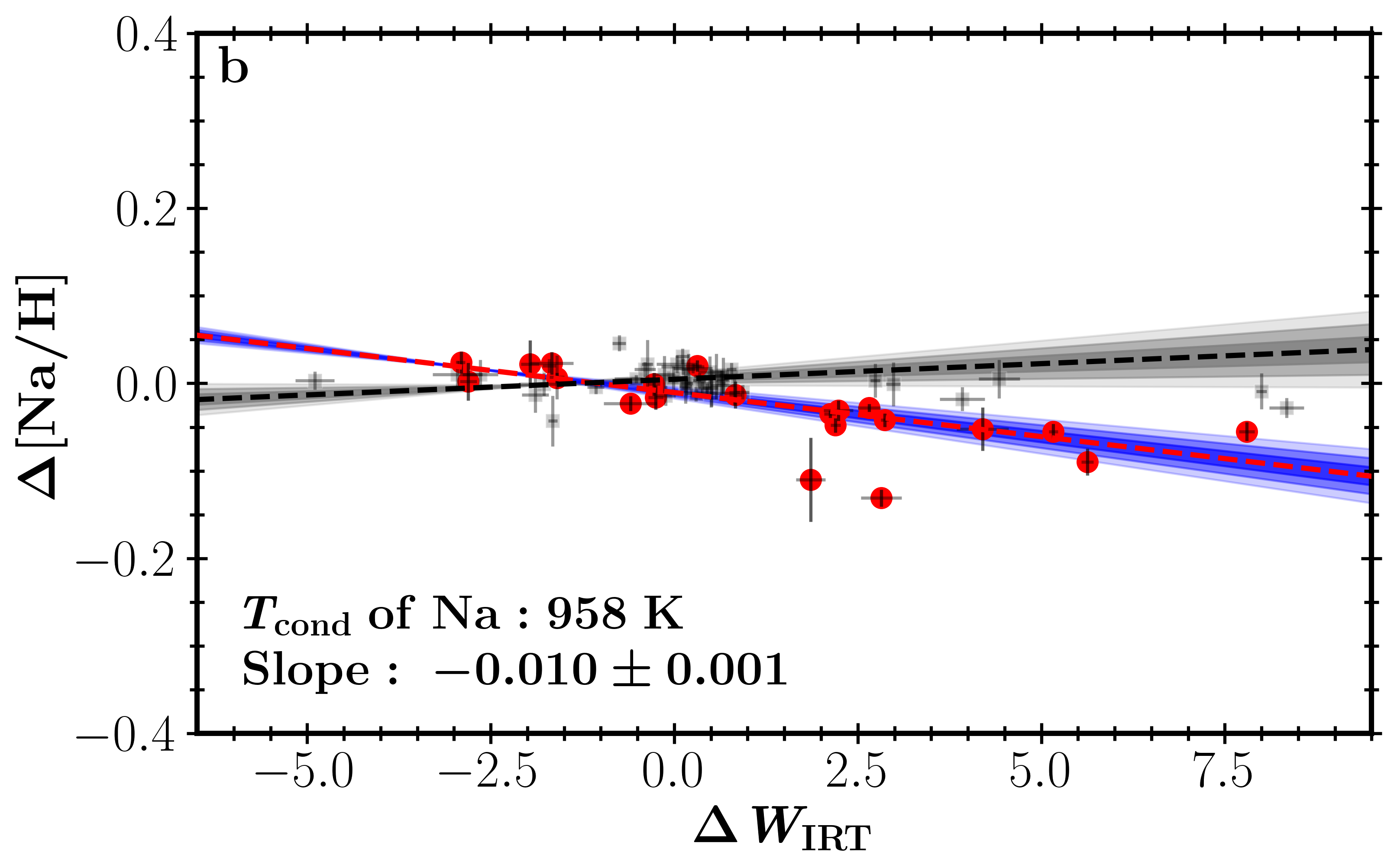}}\\
\resizebox{\columnwidth}{!}{\includegraphics[trim=0cm 1.96cm 0cm 0cm, clip]{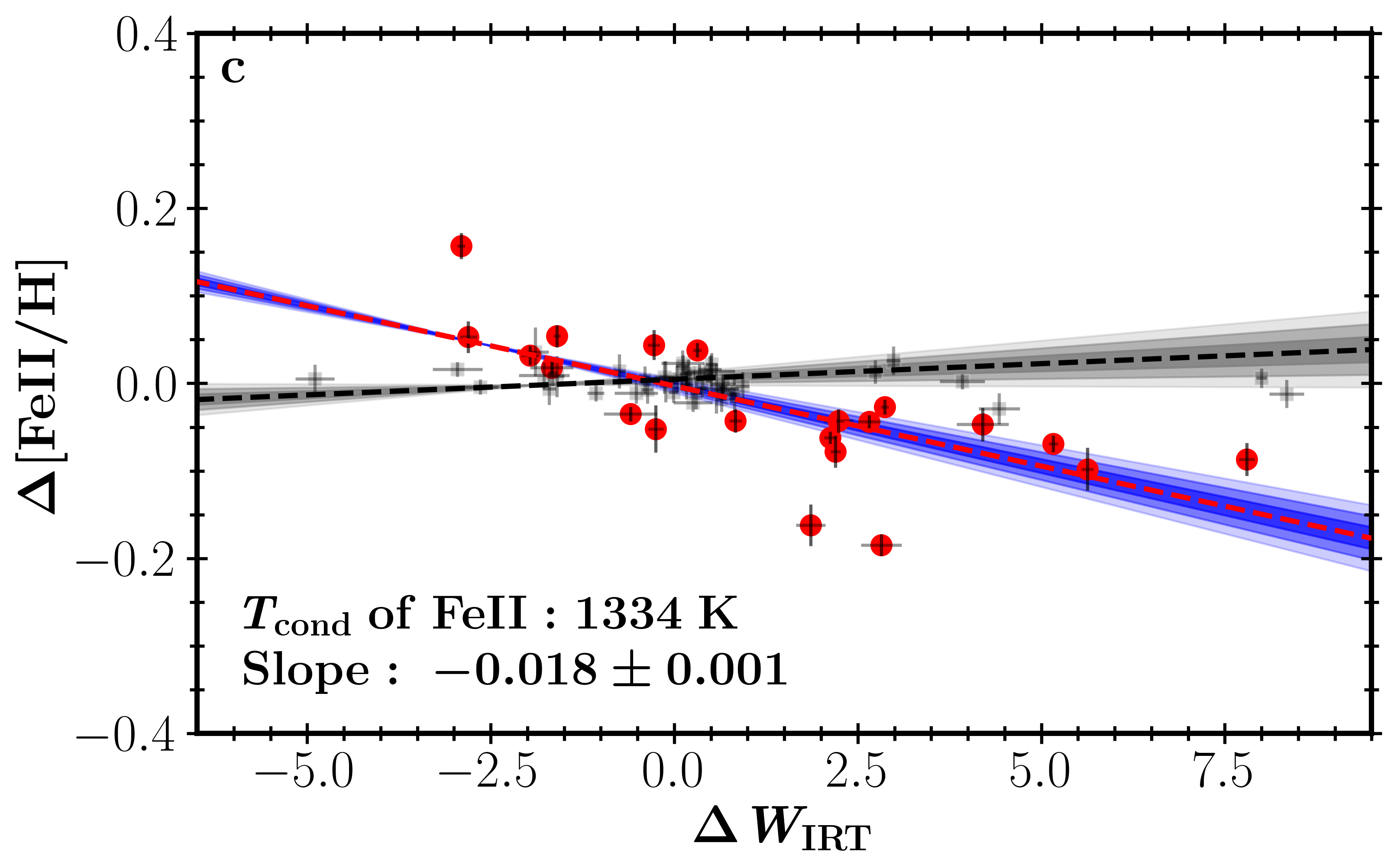}}\\
\resizebox{\columnwidth}{!}{\includegraphics{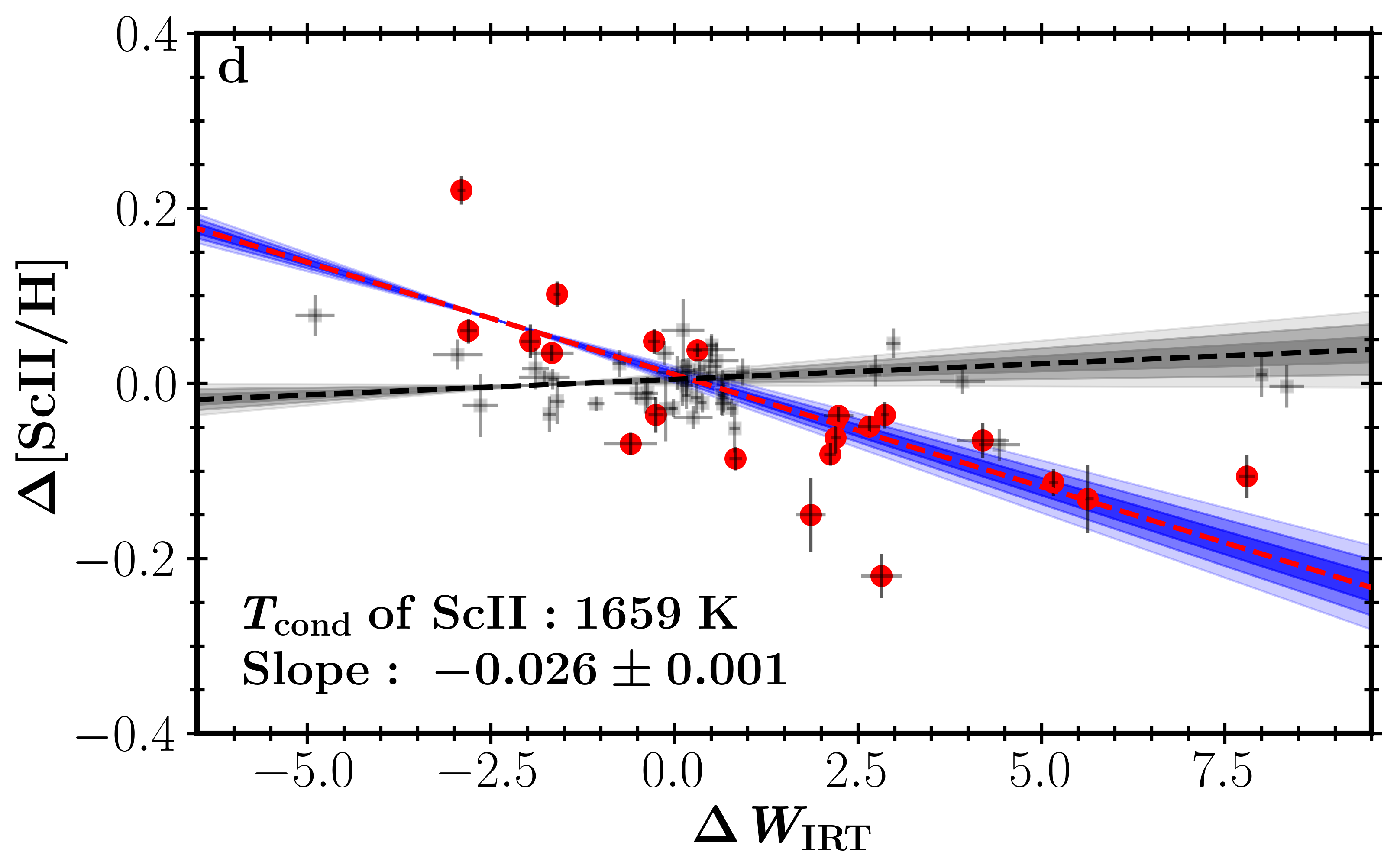}}\\
\caption{\bc{Relationships between the differential activity index, \dwirt, and the elemental abundance difference, $\Delta$[X/H], for Carbon (\textbf{a}), Sodium (\textbf{b}), Iron II (\textbf{c}), and Scandium II (\textbf{d}). The condensation temperatures of these elements increase from 40 K for C to 958 K for Na, 1334 K for Fe II, and 1659 K for Sc II, as noted in each panel. Red circles indicate star pairs with chemical anomalies, while grey circles represent those without. The red dashed line shows the best linear fit to the \dwirt–$\Delta$[X/H] relation for the chemical anomalies sample, whereas the black dashed line represents the fit for the chemical homogeneous sample. Shaded regions indicate 1$\sigma$, 2$\sigma$, and 3$\sigma$ uncertainties, with blue for the chemical anomalies sample and grey for the homogeneous sample (see legend in panel a). The slope of the red dashed line and its 1$\sigma$ uncertainty are indicated in each panel. }
}
\label{fig:ActivityAbundance}
\end{center}
\end{figure}

\section{Differential activity correlates with elemental abundance difference}
In Fig.~\ref{fig:chemistrytcondactivity}, we present the relationship between the differential activity index, \dwirt, and the slope of the condensation temperature versus elemental abundance difference, \dxhdtcont (the same slope seen in Fig.~\ref{fig:xhtcond}), for 21 co-natal star pairs with chemical anomalies and available \dwirt\ measurements. The other co-natal pairs, which do not display chemical anomalies, have \dxhdtcont\ values consistent with zero by definition and are therefore not shown. This figure reveals a correlation between \dwirt\ and \dxhdtcont\ in these pairs, as indicated by the best-fitting linear model.  Specifically, pairs with high |\dxhdtcont| values tend to exhibit strong differential activity.
Conversely, four pairs with lower \dxhdtcont\ values ($-0.3 <$ \dxhdtcont $< 0.3$) also show low differential activity ($-0.7 <$ \dwirt $< 0.7$), suggesting minimal or no exoplanetary impact. 
\bc{Among the 21 systems, four pairs identified by \citet{liu2024} as potential cases of exoplanet engulfment (marked with green arrows), where observed elemental abundance differences match theoretical models of planetary engulfment, align with this trend.}

Fig.~\ref{fig:ActivityAbundance} shows a more detailed analysis in the relation between \dxh\ and \dwirt\ for four elements, namely Carbon (C, volatile), Sodium (Na, intermediate refractory), Iron II (Fe II, refractory), and Scandium II (Sc II, refractory). The condensation temperatures of these elements range from 40 K to 1659 K, with their values annotated in Fig.~\ref{fig:ActivityAbundance}. While no clear correlation is observed for C among the pairs with chemical anomalies, an \textit{inverse} correlation is evident for Na, with a slope of $-0.010 \pm 0.001$. This inverse correlation strengthens for Fe II, with a slope of $-0.018 \pm 0.001$, and is most pronounced for Sc II, with a slope of $-0.026 \pm 0.002$. These progressively stronger inverse correlations suggest a dependence of the \dwirt–\dxh\ slope on condensation temperature for the chemical anomaly sample.

\begin{figure}
\begin{center}
\resizebox{\columnwidth}{!}{\includegraphics[trim=0cm 1.96cm 0cm 0cm, clip]{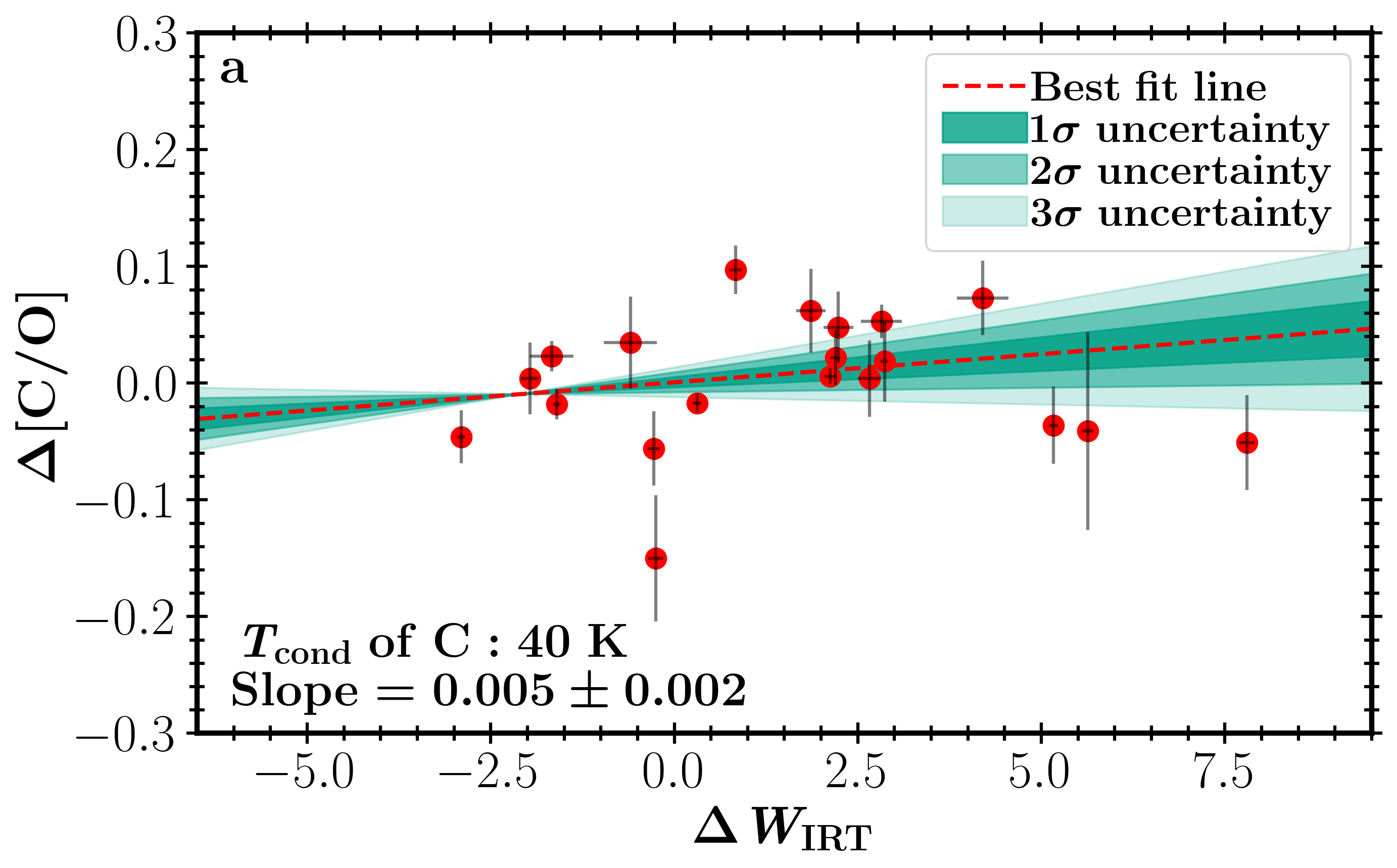}}\\
\resizebox{\columnwidth}{!}{\includegraphics[trim=0cm 1.96cm 0cm 0cm, clip]{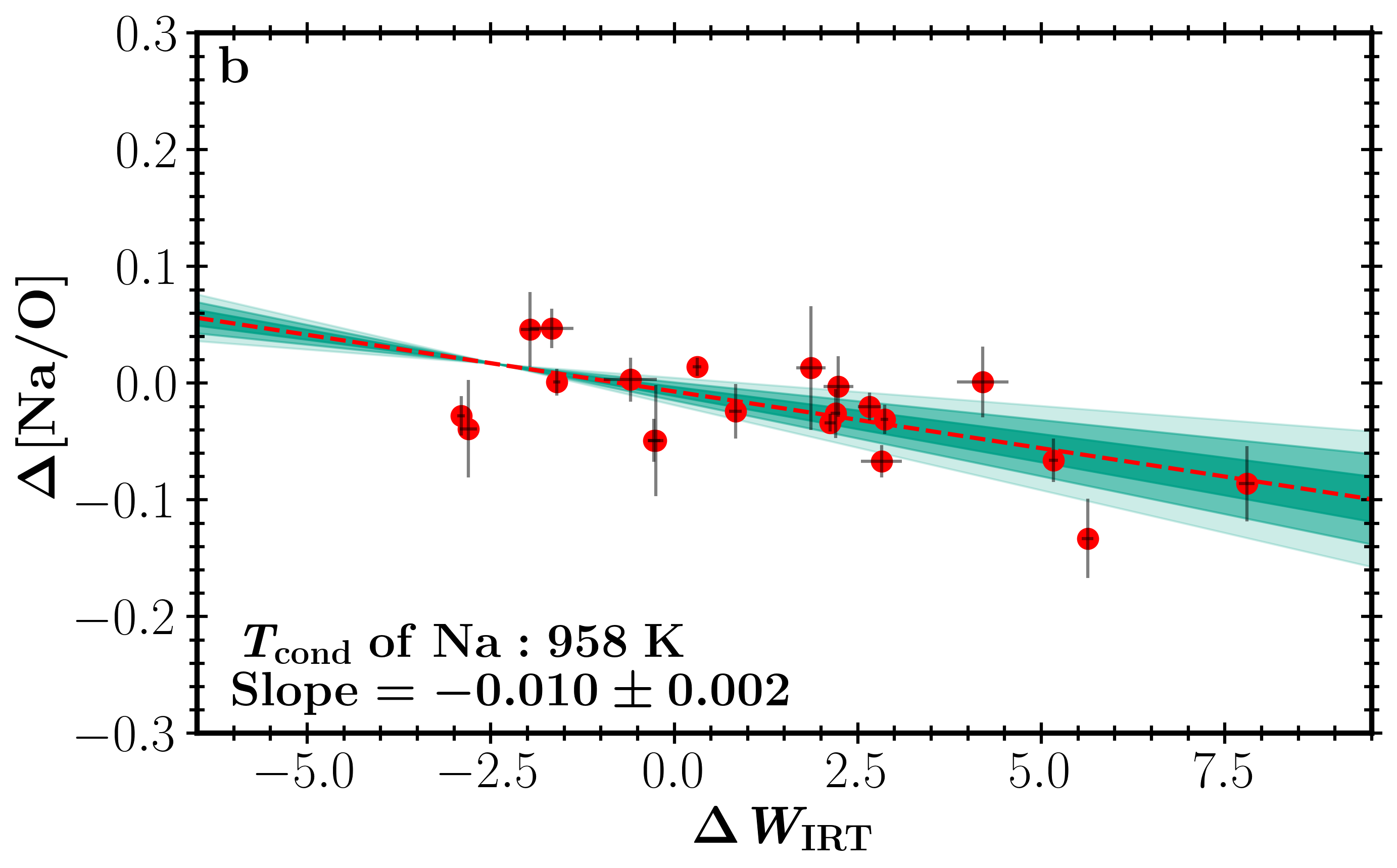}}\\
\resizebox{\columnwidth}{!}{\includegraphics[trim=0cm 1.96cm 0cm 0cm, clip]{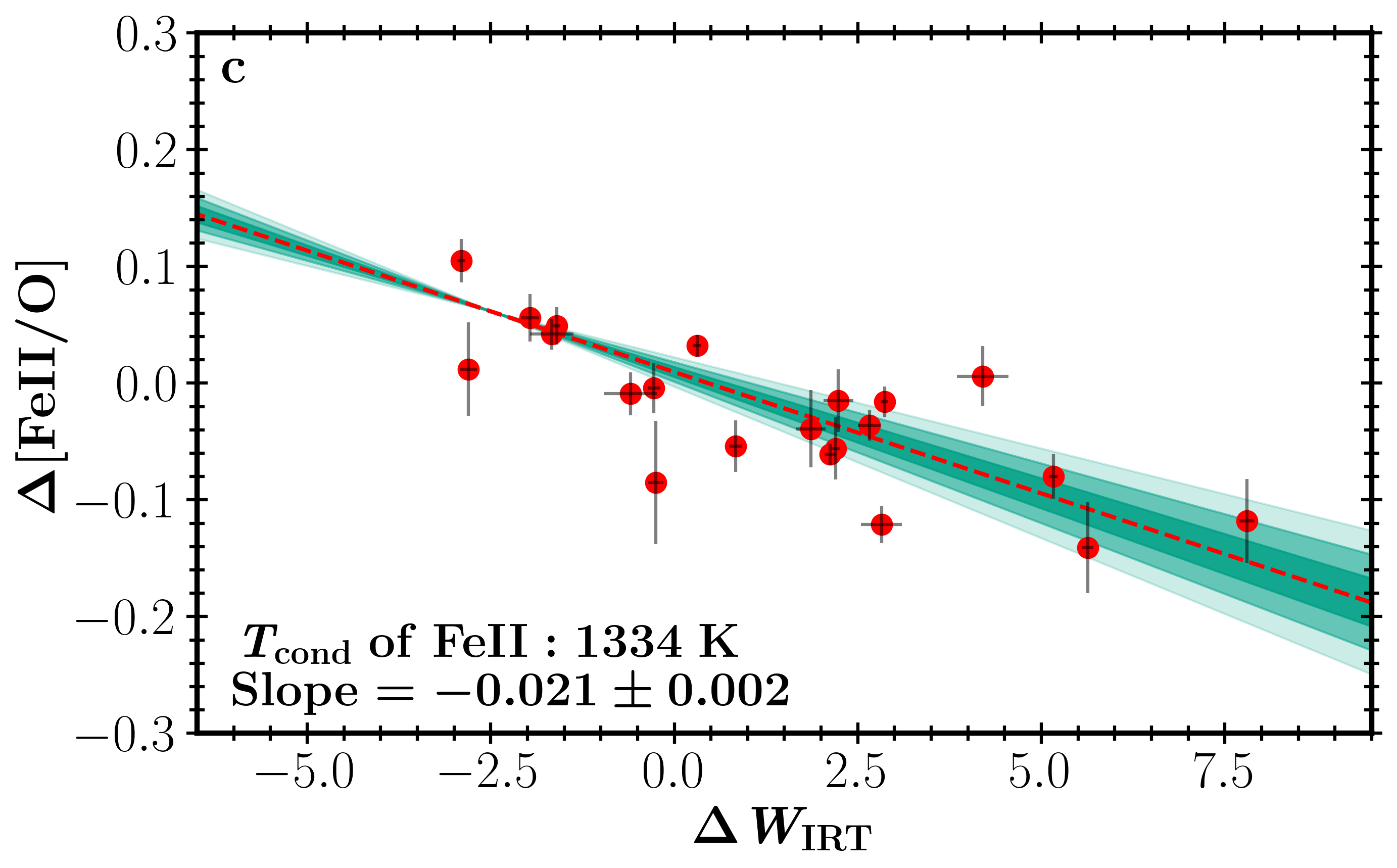}}\\
\resizebox{\columnwidth}{!}{\includegraphics{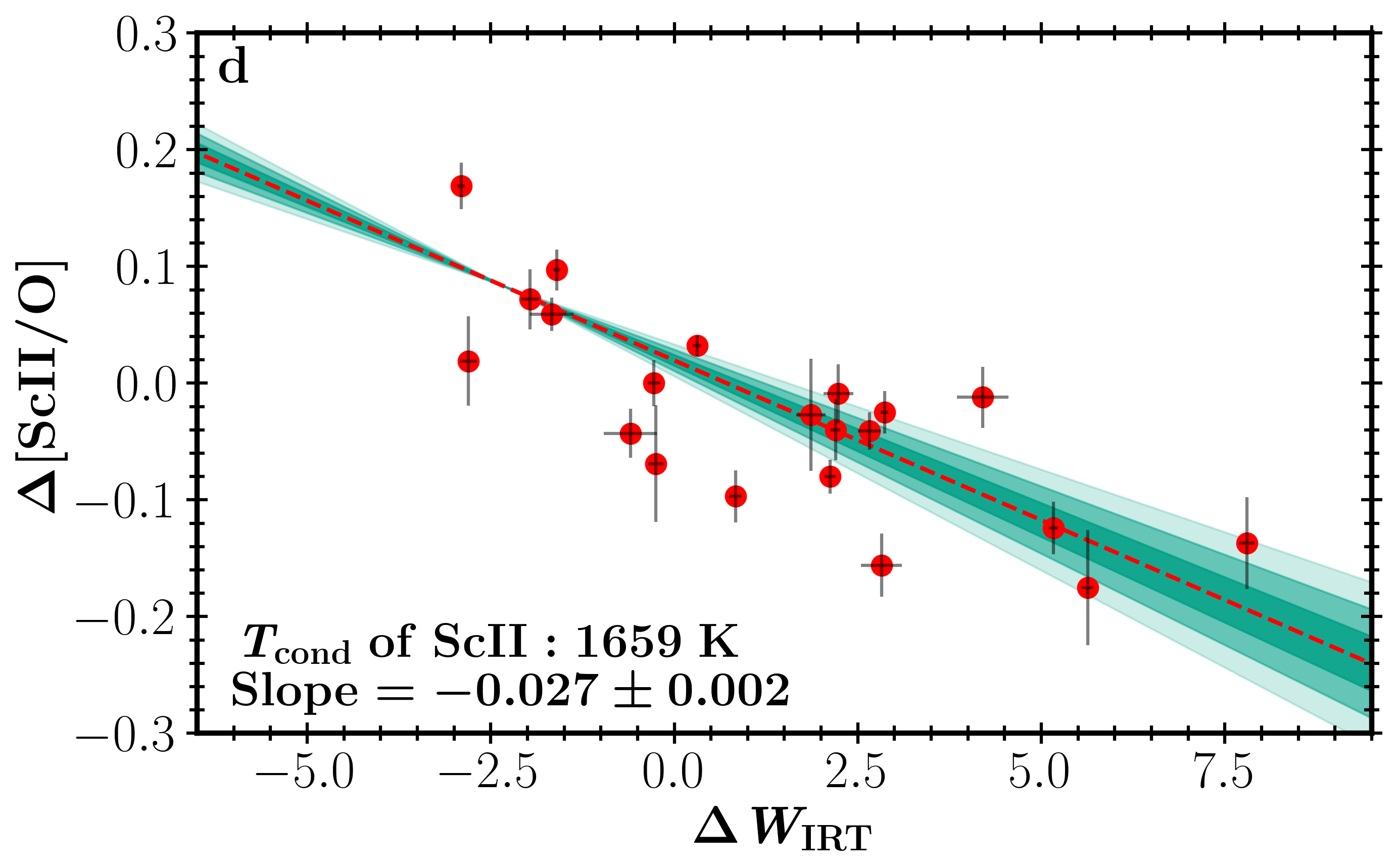}}\\
\caption{Similar to Fig.~\ref{fig:ActivityAbundance}, but now replacing \dxh\ with another metric on the vertical axes, \dxo, the differential abundance ratio between element $X$ and element Oxygen (O). Here, $X$ represents C, Na, Fe II and Sc II, as in Fig.~\ref{fig:ActivityAbundance}. We note that in addition to \dxhdtcont, \dxo\ can serve as an additional indicator of exoplanetary signatures, given that refractory elements rather than volatile elements are potentially effective tracers of exoplanetary signals \citep[e.g.][]{chambers2010, booth2020}.}
\label{fig:Activityratios}
\end{center}
\end{figure}

How does the differential activity of star pairs with homogeneous chemistry (i.e., $|\dfeh| < 3\sigma_{\dfeh}$) compare to those with chemical anomalies? In \bc{Fig.~\ref{fig:ActivityAbundance}}, we show the differential activity for systems with homogeneous chemistry, represented by grey squares. \bc{As expected, these systems show no correlation between \dxh\ and \dwirt\ due to their chemical homogeneity.} Notably, differential activity among chemically homogeneous samples is concentrated around zero, while that of chemically anomalous samples (red circles) is more broadly and evenly distributed. Quantitatively, approximately 78\% of the chemically homogeneous pairs have \dwirt\ values between -2 and 2, with a standard deviation of 0.8. For the remaining 22\%, this increased differential activity may be due to phase differences in short-term (stellar rotation) and long-term (activity cycles) activity variations, even though stars within each pair generally have similar stellar properties, suggesting comparable activity levels. 

We also perform a two-sample Kolmogorov-Smirnov (KS) test to evaluate whether the two \dwirt\ distributions shown in \bc{Fig.~\ref{fig:ActivityAbundance}} are statistically similar. \bc{As the same sample of 21 stars is analyzed for each element, the \dwirt\ distributions are identical across the four elements shown, except for carbon (top panel), where the sample consists of 20 stars}. The test yields a p-value of 0.008, indicating a rejection of the null hypothesis at the 1\% significance level and confirming that the two \dwirt\ distributions are significantly different. Overall, \bc{Fig.~\ref{fig:ActivityAbundance}} suggests that pairs without chemical anomalies tend to exhibit weaker differential activity, consistent with our finding that chemically anomalous pairs with potential planetary influences show stronger differential activity (see Fig.~\ref{fig:chemistrytcondactivity}). 

After examining elemental abundance differences, we now explore pairs with chemical anomalies using another metric, the differential abundance ratio, $\Delta$[$X$/O] = [$X$/O]$_{\rm obj}$ - [$X$/O]$_{\rm ref}$. Here, the differential represents the difference between the object and reference stars, and the abundance ratio is defined between element $X$ and Oxygen (O). We focus on $\Delta$[$X$/O] ratios, particularly $\Delta$[refractory/volatile], as they provide a direct comparison of elements that may be influenced differently by planetary processes, with refractory elements being more affected and volatile elements largely unaffected \citep[e.g.,][]{chambers2010, booth2020}. Similar to Fig.~\ref{fig:ActivityAbundance}, we again analyse $X$ = C, Na, Fe II, and Sc II in Fig.~\ref{fig:Activityratios}. We find that $\Delta$[C/O], which represents the ratio between two volatile elements (C and O), shows an insignificant correlation with differential activity. In contrast, the correlation becomes progressively stronger from $\Delta$[Na/O] to $\Delta$[Fe II/O] and $\Delta$[Sc II/O]. 

\begin{figure*}\includegraphics[width=0.8\textwidth]{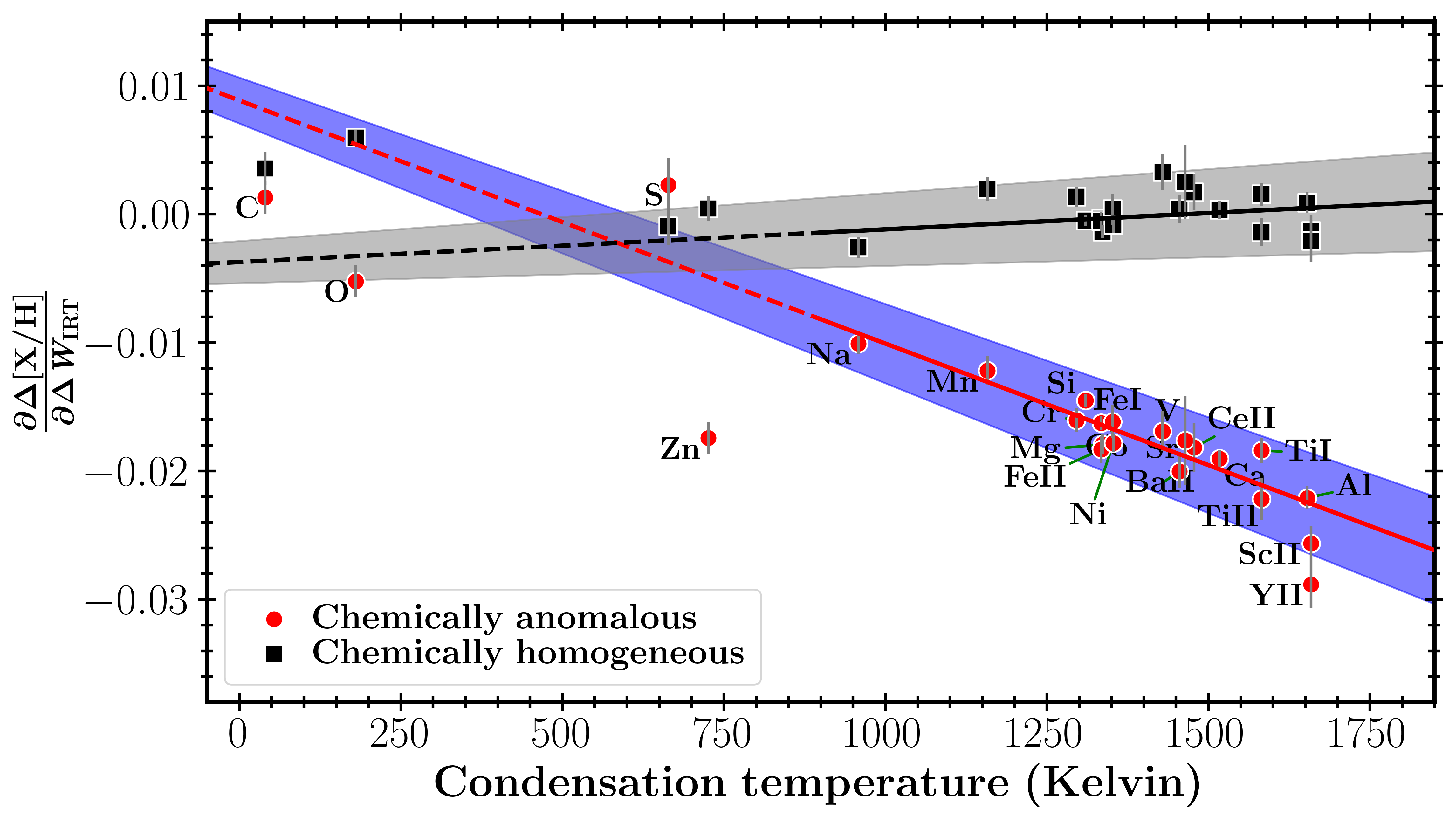}
\caption{Correlation between condensation temperature and the slope of the \dwirt–\dxh\ relation (i.e., \dxhdwirt, with 4 examples illustrated in Fig.~\ref{fig:ActivityAbundance}), where \textit{X} represents 21 individual elements annotated in the plot. Co-natal systems with chemical anomalies (i.e., $|\dfeh| \geq 3\sigma_{\dfeh}$) are represented by red circles, while those without chemical anomalies (i.e., $|\dfeh| < 3\sigma_{\dfeh}$) are represented by black squares (see legend). The red and black solid lines represent the best linear fits for the chemically anomalous and homogeneous groups, respectively, considering only refractory elements with condensation temperatures greater than $900~\mathrm{K}$. These lines are extrapolated to lower condensation temperatures ($<900~\mathrm{K}$) and are shown as dashed lines in this range. The blue and grey shaded regions represent the 1--$\sigma$ uncertainties of the fits.  The slope of the red dashed line is $\left(-1.89\pm0.13\right) \times 10^{-5}$, while the slope of the black dashed line is $\left(2.54\pm1.19\right) \times 10^{-6}$. Including volatile elements in the fitting slightly alters the slopes and  intercepts.}\label{fig:CondenTeff}
\end{figure*}

Expanding the correlation analysis to all 21 elements, Fig.~\ref{fig:CondenTeff} illustrates the significance of the inverse relationship between \dxh\ and \dwirt\ (i.e., \dxhdwirt, represented as the slopes of the dashed lines shown in Fig.~\ref{fig:ActivityAbundance}) as a function of condensation temperature. Systems with chemical anomalies (red circles) exhibit stronger anti-correlations for elements with higher condensation temperatures. That is to say, these anti-correlations are more pronounced for refractory elements (e.g., Ti and Al) compared to volatile elements (e.g., C and O). In contrast, the star pairs without chemical anomalies (black squares) show no strong correlations. Quantitatively, 
the slope of the red dashed line is $\left(-1.89\pm0.13\right) \times 10^{-5}$, while the slope of the black dashed line is $\left(2.54\pm1.19\right) \times 10^{-6}$,
indicating that the correlation for systems with anomalies is much stronger. 

\bc{Intriguingly, Earth-like elements such as C, O, Mg, Si, Ca, Ti, Cr, and Fe \citep[e.g.,][]{xu2014, harrison2021}, whose abundances are available in this work, follow the overall trend (see Fig.~\ref{fig:ActivityAbundance}). This suggests that rocky planets or giant planets with rocky cores may be linked to the observed chemical anomalies.}

\begin{figure*}
\includegraphics[width=0.8\textwidth]{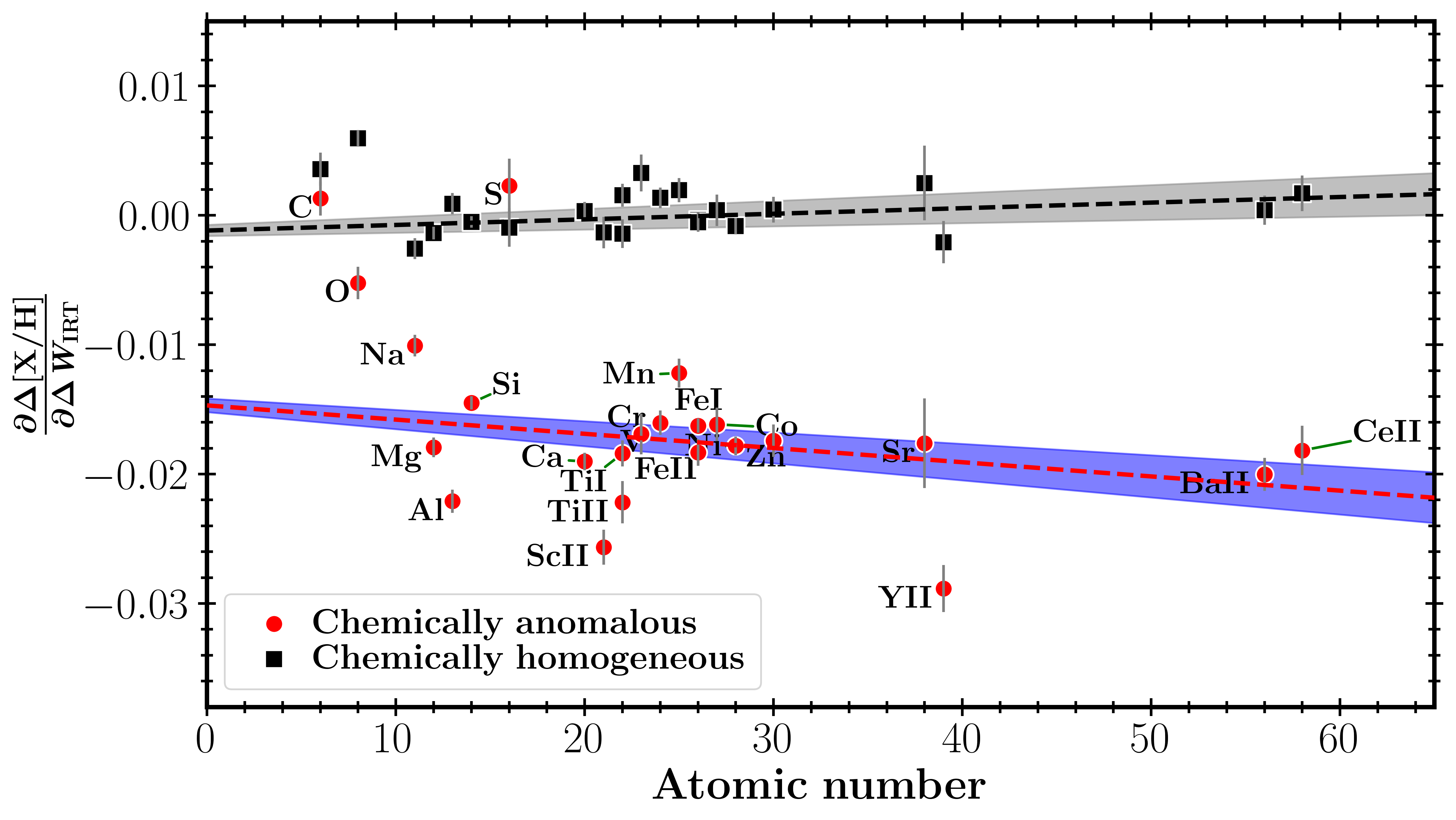}
\caption{Similar to Fig.~\ref{fig:CondenTeff} but for highlighting the correlation between atomic number and the slope of the \dwirt\---$\Delta$[X/H] relation (i.e. \dxhdtcont) for the chemically anomalous (red circles) and homogeneous (black squares) groups. The red and black dashed lines are the best fits considering only refractory elements, similar to Fig.~\ref{fig:CondenTeff}. The slope of the red dashed line is $\left(-1.10\pm0.22\right) \times 10^{-4}$, while the slope of the black dashed line is $\left(4.32\pm0.18\right) \times 10^{-5}$. }\label{fig:atomicnumber}
\end{figure*}

\begin{figure*}
\begin{center}
\resizebox{0.48\textwidth}{!}{\includegraphics{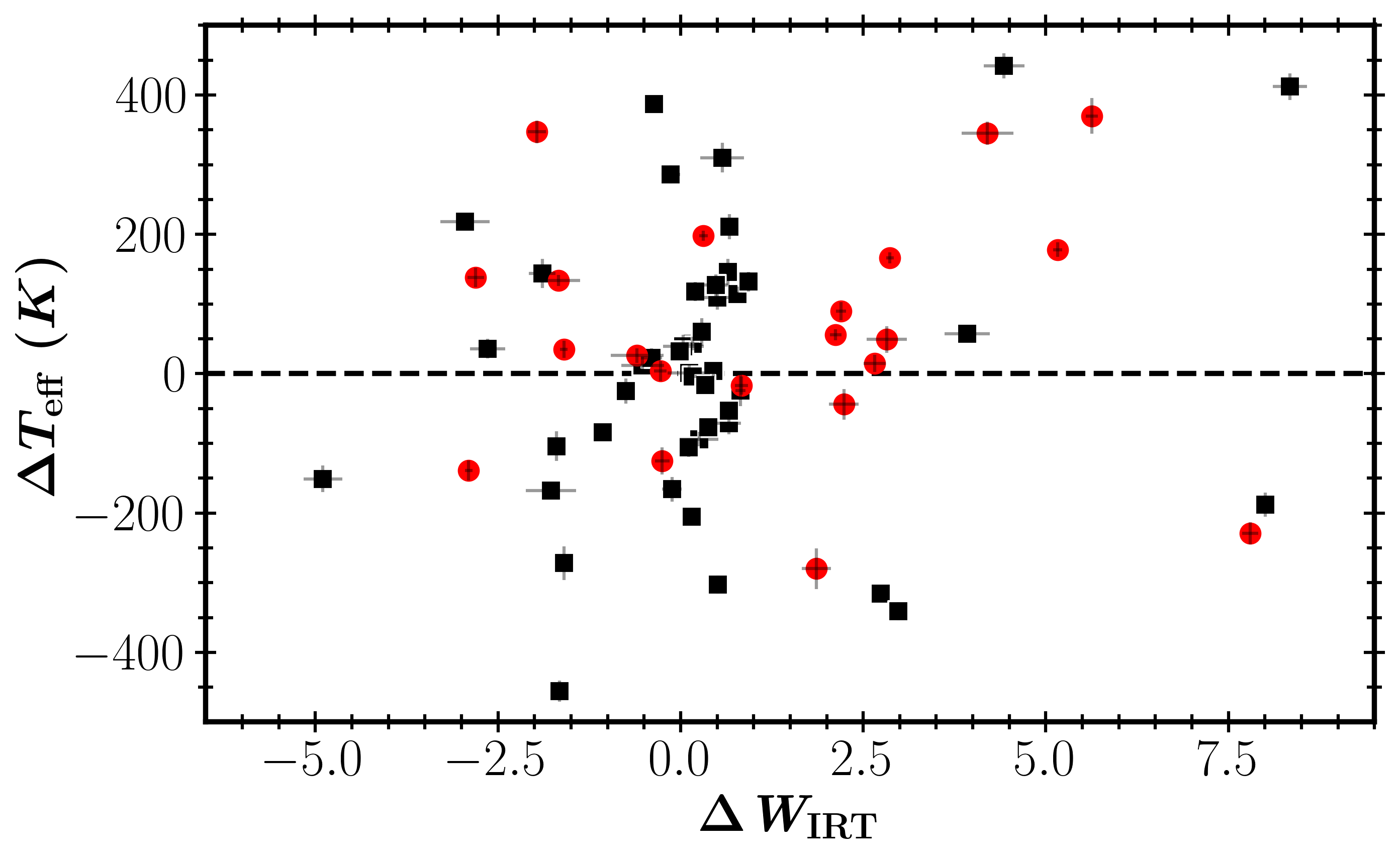}}
\resizebox{0.48\textwidth}{!}{\includegraphics{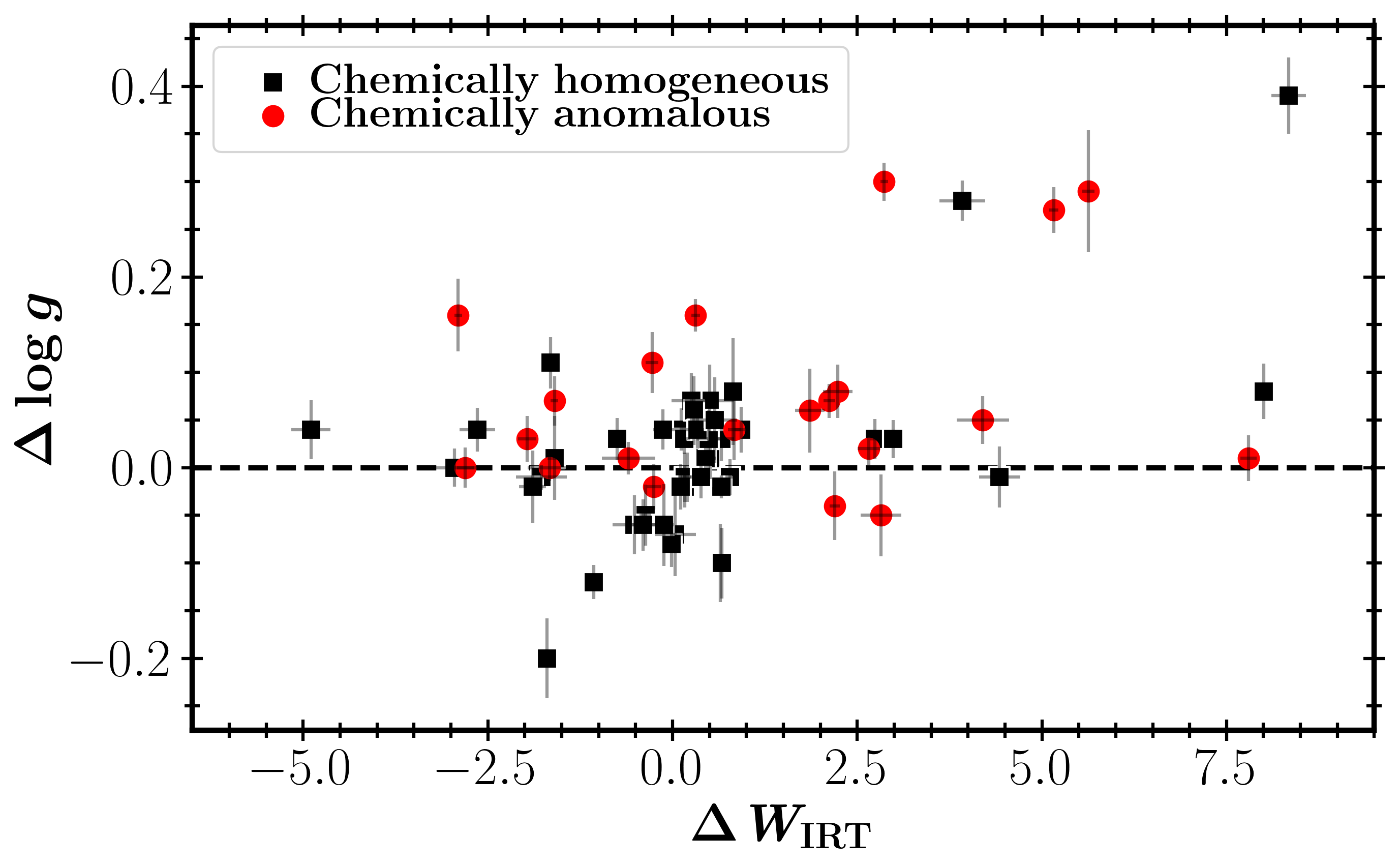}}\\
\caption{Differences in effective temperature ($\Delta \teff$, left panel) and surface gravity ($\Delta \logg$, right panel) between the two stars in a pair as a function of their activity difference (\dwirt). Red circles represent the 21 chemically anomalous pairs ($|\dfeh| \geq 3\sigmafeh$), while black squares indicate the 46 chemically homogeneous pairs ($|\dfeh| < 3\sigmafeh$). No clear correlation is observed between $\Delta \teff$ and \dwirt\ or between $\Delta \logg$ and \dwirt\ in either sample.}
\label{fig:delta_teff_logg}
\end{center}
\end{figure*}

Is the correlation with condensation temperature shown in Fig.~\ref{fig:CondenTeff} merely a variant of the correlation with atomic number, given that refractory elements are generally heavier than volatile elements? To check this, in Fig.~\ref{fig:atomicnumber}, we examine the significance of the correlation between \dxhdwirt\ and atomic number and calculate the slope of the red dashed line to be 
$\left(-1.10\pm0.22\right) \times 10^{-4}$.
To properly compare this value with that of the red dashed line in Fig.~\ref{fig:CondenTeff}, we rescaled the condensation temperature by dividing it by a factor of 28.6 (the maximum condensation temperature divided by the maximum atomic number) to match the scale of atomic number. After this adjustment, we find that the slope of the correlation between \dxhdwirt\ and the rescaled atomic number is one-fifth of the slope of the correlation shown in Fig.~\ref{fig:CondenTeff}. Additionally, the scatter (standard deviation) of the red circles around the best fit, represented by the red solid line in Fig.\ref{fig:atomicnumber}, is 0.004, which is twice as large as the scatter shown in Fig.\ref{fig:CondenTeff}. 
Therefore, the weaker and less tightly constrained correlation with atomic number compared to condensation temperature ( Fig.\ref{fig:CondenTeff} vs. Fig.\ref{fig:atomicnumber}) suggests that condensation temperature is a stronger correlate of the observed abundance trends.

It is noteworthy that there is no clear correlation between \dwirt\ and $\Delta \teff$ or between \dwirt\ and $\Delta \logg$ (see Fig.~\ref{fig:delta_teff_logg}), indicating that these observed trends are not driven by subtle differences in stellar structure among the star pairs. To quantify this, we perform Kendall’s Tau test, a non-parametric statistical method for assessing the correlation between two variables, with the null hypothesis being the absence of correlation. The test yields p-values of 0.571 for \dwirt\ vs. $\Delta \teff$ and 0.432 for \dwirt\ vs. $\Delta \logg$. Since these p-values exceed the commonly used significance thresholds of 0.01 and 0.05, the results indicate that we fail to reject the null hypothesis, concluding that there is no significant correlation in either case. Additionally, the inverse relationships between \dxh\ and \dwirt\ remain significant when we restrict the analysis to star pairs with solar-like \teff\ in the range $5572~\text{K} < \teff < 5972~$K.

\section{Discussions}\label{sect: discussion}
\subsection{Possible planet imprints suggested by differential activity}
\citet{liu2024} analyzed elemental abundances in the C3PO sample of co-moving stellar pairs and found that at least 8\% of the pairs exhibit chemical anomalies. Their Bayesian analysis demonstrated consistency between the observed abundance differences and theoretical models of planetary engulfment, suggesting that planetary ingestion offers a better explanation for these anomalies compared to models of random abundance scatter or atomic diffusion effects.

In this work, we focus on differential magnetic activity directly calculated from stellar spectra. The \textit{inverse} correlation between magnetic activity and chemical anomalies revealed in this study provides new insights into this phenomenon. While only 8 of the 21 chemically anomalous pairs show strong Bayesian evidence favoring planet engulfment models in \citet{liu2024}, we chose to analyze the complete set of chemically anomalous pairs for several reasons. 

First, the conservative Bayesian evidence threshold used in \citet{liu2024} was specifically calibrated through mock simulations to identify chemical signatures consistent with planet formation or ingestion of Earth-like composition planets. The choice of Earth-like composition, while somewhat arbitrary, reflects the limitation that Earth remains the only planet whose detailed chemical composition is well understood. Therefore, other pairs could still harbor planet-related signatures, albeit with different compositional patterns.

Second, since the C3PO program targets stellar twins specifically to mitigate stellar evolution effects, and given that ISM inhomogeneity is unlikely at scales below $10^6$ AU (the separation cutoff in this study), planet-related processes remain among the few viable explanations for the observed chemical anomalies. Importantly, our key findings, such as the trend shown in Fig.~\ref{fig:chemistrytcondactivity}, remain robust even when considering only the pairs with high Bayesian evidence.

According to \citet{guo2023}, stars that experience planetary engulfment exhibit shorter rotation periods compared to those without planets due to angular momentum transfer, with this difference increasing over time. While these stars maintain shorter rotation periods post-engulfment, for solar-mass host stars—such as those in our sample (median mass of 1.05~\msun)—this rotational difference rapidly diminishes, independent of the planet's mass or the star's initial rotation period. Given the well-established correlation between rotation rate and stellar activity \citep[e.g.,][]{isik2023}, planetary engulfment scenarios would predict either a weak or positive correlation between \dxh\ and \dwirt. However, our observation of a strong negative correlation appears inconsistent with this expectation.

On the other hand, exoplanet formation processes may deplete elemental abundances in host stars, as refractory elements can become sequestered during the formation of terrestrial planets, reducing their presence in the stellar atmosphere \citep[e.g.,][]{chambers2010}. Furthermore, close-in planets could enhance host star activity through tidal and magnetic interactions \citep[][also see the introduction for more references]{poppenhaeger2014, kavanagh2021}. This scenario predicts a negative correlation between \dwirt\ and \dxh, both of which are calculated using the same designation of object and reference stars within each pair, aligning with our observations.

Another mechanism in the exoplanet formation scenario that could enhance stellar activity involves shortened circumstellar disk lifetimes. Planets form within these disks during the first few million years of stellar evolution. During this early stage, the central star typically remains magnetically coupled to the disk through ``disk locking," preventing stellar spin-up through accretion and contraction toward the main sequence \citep{koenigl1991, colliercameron1993, edwards1993, shu1994}. 
\bc{However, \citet{booth2020} suggested that giant planets like Jupiter can create a gap in the gas disk, forming a pressure trap that prevents refractory-rich dust from accreting onto the host star. As a result, this gap can lead to the depletion of refractory elements in the star.  Meanwhile, we expect that this gap decouples the outer disk from the host star, potentially enhancing stellar contraction during the pre-main sequence (PMS) phase and shortening the disk’s lifetime. This aligns with the findings of \citet{rosotti2015}, who showed that giant planets create gas gaps that reduce the accretion rate, leading to shorter disk lifetimes. Furthermore, theoretical studies by \citet{gallet2015} and  \citet{roquette2021} demonstrate that shorter disk lifetimes are associated with faster stellar rotation. Consequently, the host star may spin up due to this enhanced contraction, resulting in increased rotation and activity during the main sequence \citep[e.g.,][]{monsch2023}.}

\begin{figure*}
\begin{center}
\resizebox{\textwidth}{!}{\includegraphics{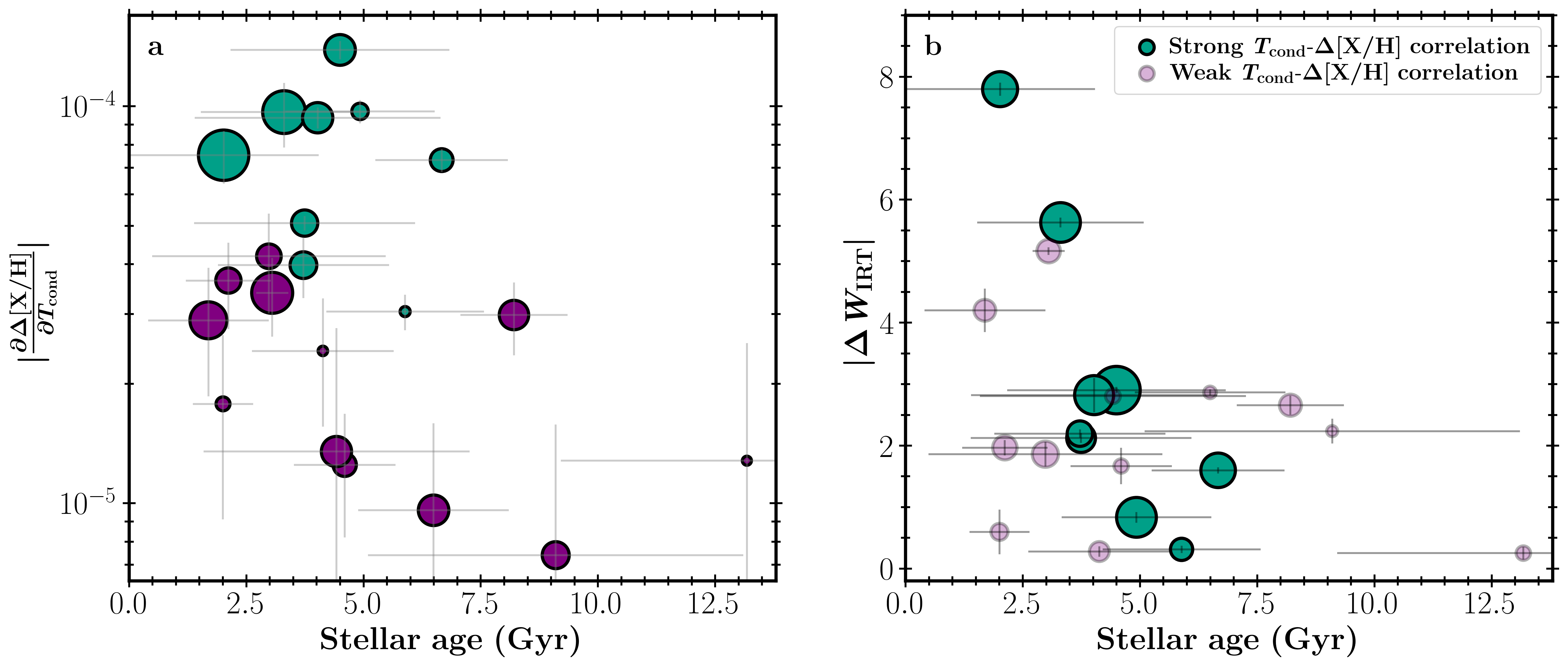}}
\caption{\textbf{Panel a:} Relationship between stellar age and the absolute slope of the $T_{\rm{cond}}$–\dxh\ correlation for 21 co-natal systems exhibiting chemical anomalies. The symbol size is proportional to the absolute value of differential activity, $|\dwirt|$. Stellar ages are derived from isochrone fitting based on \teff, \logg, and \feh\ measurements (see Section~\ref{sample}). Cyan symbols represent systems with a strong correlation between $T_{\rm{cond}}$ and \dxh ($5\sigma$ level with $\sigma$  being the uncertainty of the slope), while purple symbols indicate systems without such significant correlations. \textbf{Panel b:} Absolute differential activity ($|\dwirt|$) as a function of stellar age. The colour scheme follows that of Panel a, except for lighter purple to highlight systems with strong $T_{\rm{cond}}$–\dxh\ correlations (shown in cyan). The symbol size is proportional to |\dxhdtcont|.
}
\label{fig:ageeffect}
\end{center}
\end{figure*}

Given these considerations, we can hypothesize that signatures of exoplanet formation gradually weaken during a star's main sequence lifetime. As shown in Fig.~\ref{fig:ageeffect}, systems with strong ($5\sigma$, where $\sigma$ is the uncertainty in \dxhdtcont) correlations between $T_{\rm{cond}}$ and \dxh\ (cyan circles, 45\% of systems with chemical anomalies) exhibit a trend between stellar age and absolute differential activity, where younger stars show stronger activity differences. In contrast, systems without strong $T_{\rm{cond}}$–\dxh\ correlations (dark purple circles in Fig.~\ref{fig:ageeffect}a) display a much weaker trend between \dwirt\ and age (light purple circles in Fig.~\ref{fig:ageeffect}b). These results hint at a connection between stellar age and the longevity of planetary chemical signatures. Given the relatively large uncertainties (1.6 Gyr) in age measurements, future studies focusing on acquiring additional data and improving age precision will be valuable for confirming and further refining these trends.

If the $T_{\rm{cond}}$ signal is indeed due to planetary signatures—an assumption that warrants further verification by ruling out other possibilities (see next section for a discussion)—then planet formation, rather than planet ingestion, appears more likely to create both the observed chemical patterns (especially those with high Bayesian evidence) and the negative correlation with magnetic activity. 

From a theoretical perspective, \citet{booth2020} argued that the 0.04 dex depletion in refractories observed in the Sun relative to solar twins due to planet formation can be readily explained by factors such as large disc masses (e.g., \(M_D \approx 0.2 \, M_\odot\)), low photoevaporation rates, the shrinking convective zone after 2 Myr, and earlier planet formation, which collectively enhance dust depletion in the system. However, recent simulations in \citet{huhn2023} suggest that planet formation should not produce signals of the magnitude observed in C3PO (out of 21 pairs, 2 with \(\Delta\left[\text{Na/H}\right]\) and 7 with \(\Delta\left[\text{ScII/H}\right]\) values ranging from 0.1 to 0.2 dex). Moreover, if the chemical patterns and rotation-induced magnetic activity arise from planet formation, maintaining such signals up to 4 Gyr requires further theoretical studies.

Resolving this theoretical discrepancy is beyond the scope of this observational paper, which aims to report the correlation between magnetic activity and chemical abundance patterns rather than arrive at theoretical explanations. Nevertheless, if planet-induced signals can persist for several Gyr, this may help explain the long-observed chemical anomalies of the Sun compared to other solar analogs \citep{bedell2018}.

\subsection{Alternative Explanations}
Here we examine potential alternative mechanisms that could produce our observed signals. First, we consider stellar activity cycles. According to \citet{spina2020}, activity cycles affect elemental abundance measurements through magnetic field variations in the stellar atmosphere, which alter absorption line equivalent widths. While such cycle-induced variations are real and could potentially complicate abundance analyses, \citet{spina2020} demonstrated that these abundance variations show no correlation with condensation temperature. This is a crucial distinction from our observations, as the $T_{\rm{cond}}$ correlation (Fig.~\ref{fig:CondenTeff}) represents a key signature in our dataset. 
\bc{This difference suggests that the correlation we detect between abundance variations and activity is not driven by the physics of activity cycles, consistent with the idea that exoplanets may be the underlying cause.}

Internal stellar mixing processes present another possible mechanism that warrants careful consideration. In stars, atomic diffusion or gravitational settling causes heavier elements to sink while lighter ones rise \citep[see][for a recent review]{alecian2023}. Rotational mixing can counteract this process by transporting elements from the stellar interior to the surface, with faster rotation leading to more efficient mixing \citep[e.g.,][]{deal2020}. This mechanism would predict that slower-rotating stars show greater depletion of heavy elements, as reduced mixing efficiency would allow more settling to occur. However, our observations reveal the opposite trend: magnetically more active stars (which rotate faster) exhibit greater chemical depletion. Additionally, atomic diffusion effects are expected to produce systematic trends with stellar evolutionary state, particularly surface gravity. The C3PO survey's focus on stellar twins specifically minimizes such evolutionary effects, as our pairs have nearly identical surface gravities. Indeed, we find (1) a weaker and less tightly constrained correlation of differential activity  with atomic number compared to condensation temperature (Fig.~\ref{fig:CondenTeff} vs. Fig.~\ref{fig:atomicnumber}) and (2) no correlation between magnetic activity and stellar surface gravity in our sample, arguing against mixing processes as the primary driver of our observed chemical patterns. These multiple lines of evidence suggest that stellar mixing processes alone cannot explain the observed relationship between magnetic activity and chemical abundances.

Another widely discussed factor influencing abundance trends with condensation temperature is Galactic Chemical Evolution (GCE), as abundance patterns can be influenced by variations in stellar ages \citep[e.g.][]{bedell2018, sun2025a}. However, \citet{spina2016a} suggested that GCE effects could be disentangled from planetary signatures by removing the [X/Fe] versus age trends. Furthermore, other studies indicate that while GCE can account for some of the observed variation, it cannot fully explain the entire trends observed in many stars \citep{nissen2016, spina2016, bedell2018}. For example, the Sun still exhibits refractory depletion even after GCE corrections are applied \citep{nissen2016, spina2016}. Therefore, GCE is likely a contributing factor but not the primary driver of these chemical anomalies. The co-natal stars in our study further minimize the impact of GCE because of their shared formation history.

\section{Conclusions}
We investigated the differential activity of 21 co-natal star pairs with chemical anomalies, selected from the C3PO program, which provides the largest and most homogeneous sample of co-moving systems to date. Each pair consists of a reference star and a companion star with similar fundamental properties (see Section \ref{sample}). The typical (median) age of these stars is approximately 3.2 Gyr, with a standard deviation of 3.5 Gyr. Using high-resolution, high-SNR spectra from the Magellan and Keck, we derived a differential activity index, \dwirt, based on the flux difference in the Ca II infrared triplet lines between the two stars in each pair, which is validated with \dsindex\ (see Fig.~\ref{fig:SIndex} and Section \ref{med: activityindex}). Using elemental abundance measurements from \citet{liu2024}, we then examined the relationships between differential activity (\dwirt), elemental abundance differences (\dxh), and dust condensation temperature. Our main findings are:

\begin{enumerate}
    \item Pairs exhibiting strong correlations between condensation temperature and elemental abundance differences also show pronounced differential activity (Fig.~\ref{fig:chemistrytcondactivity}).
    
    \item We discover significant inverse correlations between \dwirt\ and \dxh\ for refractory elements (Na, Fe II, Sc II), indicating that chemically depleted stars are magnetically more active. No such correlation exists for volatile elements (Fig.~\ref{fig:ActivityAbundance}).
    
    \item The strength of these inverse correlations increases systematically with condensation temperature (Fig.~\ref{fig:CondenTeff}).

    \item In systems with strong correlations between condensation temperature and elemental abundance difference, younger stars exhibit larger differential activity, suggesting an age-dependent effect (Fig.~\ref{fig:ageeffect}).
    
\end{enumerate}

These patterns add a new dimension to understanding the origin of stellar chemical anomalies. If the chemical signatures are indeed planet-related, then the inverse correlation between chemical depletion and magnetic activity, coupled with the condensation temperature dependence, points toward planet formation rather than engulfment as a possible mechanism (see Section~\ref{sect: discussion}). \citet{booth2020} found that the 0.04 dex depletion observed in the Sun relative to solar twins due to planet formation can be readily explained. However, \citep{huhn2023} predict chemical signatures substantially smaller than those we observe. Understanding this discrepancy and the longevity of these signals requires further investigation.

Several follow-up studies have the potential to provide more insights. 
\bc{First, radial-velocity monitoring, including multi-epoch Gaia RVS spectra from upcoming data releases (e.g., Gaia DR4), will provide valuable data to investigate whether these bright stars host giant planets. This is particularly relevant for pairs with significant differential abundances and activity (Fig.~\ref{fig:ActivityAbundance}) and offers a potential test of the \citet{booth2020} hypothesis.} Second, rotation period measurements from TESS light curves \citep[e.g.,][]{colman2024} could illuminate the connection between differential activity and rotation, though careful consideration of binary contamination effects will be necessary. Finally, theoretical work is needed to reconcile the magnitude of observed chemical signatures with planet formation models. These combined efforts will advance our understanding of how planets shape the chemical and magnetic evolution of their host stars.

\section*{Acknowledgements}
 J.Y. acknowledges the Joint Research Fund in Astronomy (U2031203) under a cooperative agreement between the National Natural Science Foundation of China (NSFC) and the Chinese Academy of Sciences (CAS). Y.S.T is supported by the National Science Foundation under Grant No. AST-2406729. This paper includes data gathered through the C3PO program with the 6.5 metre Magellan Telescope located at Las Campanas Observatory, Chile. Some of the data presented herein were obtained at the W. M. Keck Observatory, which is operated as a scientific partnership among the California Institute of Technology, the University of California and the National Aeronautics and Space Administration. The observatory was made possible by the generous financial support of the W. M. Keck Foundation. Based on observations collected at the European Organisation for Astronomical Research in the Southern Hemisphere under ESO programme 108.22EC.001. This work has made use of data from the European Space Agency (ESA) mission Gaia (https://www.cosmos.esa.int/gaia), processed by the Gaia Data Processing and Analysis Consortium (DPAC, https://www.cosmos.esa.int/web/gaia/dpac/consortium). Funding for the DPAC has been provided by national institutions, in particular the institutions participating in the Gaia Multilateral Agreement.

\section*{Data Availability}
The spectra used to calculate activity indices in this work were collected with Magellan and Keck and processed by \citet{yong2023} and \citet{liu2024}. Stellar parameters and elemental abundances were adopted from \citet{liu2024}. To facilitate the reproduction of the main results presented in this work, the data on activity indices will be made available upon reasonable request to the corresponding author.

\bibliographystyle{mnras}
\bibliography{references.bib} 

\bsp 
\label{lastpage}
\end{document}